\documentclass[preprint,3p]{elsarticle}

\usepackage{lineno,hyperref}
\modulolinenumbers[5]

\journal{Elsevier}

\usepackage[usenames,dvipsnames]{xcolor}

\usepackage{graphicx}        

\usepackage{natbib}

\usepackage{amsmath}
\usepackage{amsfonts}
\usepackage{amssymb}

\usepackage{subfigure}
\usepackage{xspace}

\newcommand{\etal}{et al.\xspace}
\newcommand{\ie}{i.\,e.\xspace}
\newcommand{\eg}{e.\,g.\xspace}
\newcommand{\cf}{cf.\xspace}
\newcommand{\fp}{\,.}
\newcommand{\fk}{\,,}

\newcommand{\mrm}[1]{\mathrm{#1}} 

\newcommand{\nc}{\newcommand}
\newcommand{\rnc}{\renewcommand}
\nc{\tra}{^{\mathrm T}}
\nc{\sref}[1]{Sect.~\ref{sec:#1}}
\nc{\srefs}[1]{Sect.~\ref{sec:#1}}
\nc{\srefo}[1]{\ref{sec:#1}}
\nc{\ssref}[1]{Subsect.~\ref{sec:#1}}
\nc{\ssrefs}[1]{Subsect.~\ref{sec:#1}}
\nc{\ssrefo}[1]{\ref{sec:#1}}
\nc{\sssref}[1]{Subsubsect.~\ref{sec:#1}}
\nc{\sssrefs}[1]{Subsubsect.~\ref{sec:#1}}
\nc{\sssrefo}[1]{\ref{sec:#1}}
\nc{\eref}[1]{Eq.~\ref{eq:#1}}
\nc{\erefs}[1]{Eqs.~\ref{eq:#1}}
\nc{\erefo}[1]{\ref{eq:#1}}
\nc{\fref}[1]{Fig.~\ref{fig:#1}}
\nc{\frefs}[1]{Figs.~\ref{fig:#1}}
\nc{\frefo}[1]{\ref{fig:#1}}
\nc{\tref}[1]{Tab.~\ref{tab:#1}}
\nc{\aref}[1]{\ref{asec:#1}}
\nc{\fig}[4][tbh]{
\begin{figure}[#1]
\centering
\includegraphics[width=#4\textwidth]{figs/#2}
\caption{#3\label{fig:#2}}
\end{figure}}
\nc{\ea}[1]{
\begin{eqnarray}
#1\end{eqnarray}}
\rnc{\matrix}[2]{\left[\!\!\begin{array}{#1}
	#2\end{array}\!\!\right]}
\rnc{\vector}[1]{\matrix{c}{#1}}
\usepackage{color}
\nc{\MODIFIED}[1]{#1}
\usepackage{wasysym}
\usepackage{tikz} 
\nc{\circled}[1]{(#1)}
\nc{\pmpar}{{}_{\scriptscriptstyle (}\!\pm\!{}_{\scriptscriptstyle )}} 

\nc{\pmr}{\varepsilon_\mrm{phy}} 
\nc{\vc}{v_\mrm{c}} 

\usepackage{stfloats}
\fnbelowfloat 
\usepackage[noprefix]{nomencl}
\makenomenclature
\usepackage{cancel}
\usepackage{cases}

\usepackage{svg}
\usepackage{multirow}
\nc{\x}[1]{\mbox{#1}}
\nc{\zo}[1]{\x{\cite{#1}}}
\nc{\wrt}{w.\,r.\,t.\xspace}
\nc{\fnote}[1]{\footnote{\samepage #1}} 
\nc{\ze}[2]{\x{${#1}~{\rm{#2}}$}}       
\rnc{\epsilon}{\varepsilon}
\rnc{\rho}{\varrho}
\rnc{\phi}{\varphi}
\rnc{\arraystretch}{1.1}
\nc{\si}[1]{_{\mathrm{#1}}} 
\nc{\e}[2]{\begin{equation} #1 \label {eq:#2} \end{equation}}
\nc{\impl}{FE model (implicit)}
\nc{\expl}{FE model (explicit)}
\nc{\mlcms}{massless-boundary ROM}
\nc{\ABAQUS}{{\sf{Abaqus}}\xspace}
\nc{\MATLAB}{{\sf{MATLAB}}\xspace}
\nc{\myquote}[1]{`#1'}
\nc{\mm}{\boldsymbol}
\nc{\dd}{\mathrm{d}}
\nc{\ii}{\mathrm{i}}
\nc{\ee}{\mathrm{e}}
\nc{\Mtil}{\mm m}
\nc{\Ktil}{\mm k}
\nc{\kbb}{\mm k_{\mrm{bb}}}
\nc{\kbi}{\mm k_{\mrm{b}\eta}}
\nc{\kbitra}{\mm k\tra_{\mrm{b}\eta}}
\nc{\kii}{\mm k_{\eta\eta}}
\nc{\qq}{\mm q}
\nc{\qqb}{\mm q_{\mrm b}}
\nc{\qqi}{\mm q_{\mrm i}}
\nc{\inv}{{}^{-1}}
\nc{\eye}{\mm I}
\nc{\nmod}{N_{\mathrm{mod}}}
\nc{\diag}{\operatorname{diag}}
\nc{\IES}{IES\xspace}
\nc{\IESs}{IESs\xspace}

\makeindex             

\begin{document}


\begin{frontmatter}
\title{Computational and experimental analysis of the impact of a sphere on a beam and the resulting modal energy distribution}
\author[addressMTU]{Felix Gehr}
\author[addressILA]{Timo Theurich}
\author[addressILA]{Carlo Monjaraz-Tec}
\author[addressILA]{Johann Gross}
\author[addressMTU]{Stefan Schwarz}
\author[addressMTU]{Andreas Hartung}
\author[addressILA]{Malte Krack}

\address[addressMTU]{MTU Aero Engines AG, Dachauer Strasse 665, 80995 Munich, Germany\\ felix.gehr@mtu.de, stefan.schwarz@mtu.de, andreas.hartung@mtu.de}
\address[addressILA]{Institute of Aircraft Propulsion Systems, University of Stuttgart, Pfaffenwaldring 6, 70569 Stuttgart, Germany\\ timo.theurich@ila.uni-stuttgart.de, carlo-daniel.monjaraz-tec@ila.uni-stuttgart.de, johann.gross@ila.uni-stuttgart.de, malte.krack@ila.uni-stuttgart.de}

\begin{abstract}
We consider the common problem setting of an elastic sphere impacting on a flexible beam.
In contrast to previous studies, we analyze the modal energy distribution induced by the impact, having in mind the particular application of impact vibration absorbers.
Also, the beam is analyzed in the clamped-clamped configuration, in addition to the free-free configuration usually considered.
{
We demonstrate that the designed test rig permits to obtain well-repeatable measurements.
The measurements are confronted with predictions obtained using two different approaches, state-of-the-art Finite Element Analysis and a recently developed computational approach involving a reduce-order model.
The innovative aspect of the latter approach is to achieve a massless contact boundary using component mode synthesis, which reduces the mathematical model order and numerical oscillations.
We show that the novel computational approach reduces the numerical effort by 3-4 orders of magnitude compared to state-of-the-art Finite Element Analysis, without compromising the excellent agreement with the measurements.
}
\end{abstract}


\end{frontmatter}


\section{Introduction} \label{sec:intro}
Although the computational methods validated in this article are quite broadly applicable, an important technical motivation of the present work were Impact Absorbers.
Thus, some technical context is given in the next subsection.
Subsequently, the relevant state of knowledge on computational methods for elastodynamic contact problems is presented.
Then, previous experimental work on the considered problem setting of a beam impacted by a sphere is analyzed.
Finally, the outline of the present work is described.

\subsection*{Technical motivation of the present work: Impact Absorbers}
An Impact Absorber consists of a small mass designed to undergo collisions with the vibrating host structure.
For instance, the mass can be of spherical shape and placed inside a cavity of the host structure with limited free-play.
Two types of Impact Absorbers must be distinguished depending on their working principle, the Impact Damper and the Impact Energy Scatterer \cite{Theurich.2019}.
The \emph{Impact Damper} relies on material dissipation in the contact region.
In contrast, the \emph{Impact Energy Scatterer} (\IES) transfers energy to high-frequency modes, where the energy is dissipated on faster time scales and, by design, away from the contact region \cite{Theurich.2019}.
As this corresponds to a dispersion effect, the \IES can also be referred to as Impact Disperser.
The idea of the Impact Damper goes back to the 1930s \cite{Paget.1937}, and it is now also known as Vibro-Impact Nonlinear Energy Sink
\footnote{
\MODIFIED{
A Nonlinear Energy Sink is a light device attached in an \emph{essentially nonlinear} way to the host structure; \ie, the attachment force has (almost) no linear part \cite{vaka2008b}.
In the case of the Impact Damper, the essentially nonlinear attachment is simply realized as free-play nonlinearity.
As any Nonlinear Energy Sink, the Impact Damper (and also the \IES) has no preferential frequency and can engage into resonance with arbitrary modes of the host structure over broad frequency ranges.
This is an important advantage over linear and nonlinear tuned vibration absorbers, which are narrow-band devices.
}}
\cite{vaka2008b,Lamarque2011,Gendelman.2012}.
In contrast to the Impact Damper, the working principle of the \IES has only recently been analyzed and understood \cite{Theurich.2019}.
The reason for this is certainly that most research on Impact Absorbers has focused on single-degree-of-freedom host structures.
In this case, it can be shown that Impact Absorbers do not achieve any vibration mitigation unless the collisions are inelastic (dissipative).
Now, if one considers flexible, \eg lightweight, host structure's with many low-frequency vibration modes, the picture completely changes.
Even purely elastic collisions then induce elastic wave propagation within the host structure and thus transfer energy from the critical low-frequency modes to high-frequency modes.
Assuming the same inherent damping ratio for all modes, high-frequency modes dissipate energy more rapidly as they accumulate more vibration cycles within a given time span.
Thus, the host structure’s inherent damping (due to \eg friction joints, distributed material and aerodynamic damping) are exploited more efficiently than in the case without \IES.
In this sense, the \IES helps the structure to damp itself.
With the described working principle, the \IES does not require dissipation in the contact region.
Consequently, the \IES and its casing can be designed to avoid the concentration of plastic behavior in the contact region (inevitably associated with damage and thus reducing its service life), which is believed to be crucial from an engineering perspective \cite{Hartung.2017}.

\subsection*{Computational methods for elastodynamic contact problems}
The design of \IESs requires the simulation of impacts between absorber and host structure and the resulting modal energy distribution.
We aim for a \emph{predictive} approach relying only on geometry and material properties.
Hence, we want to avoid phenomenological contact models, such as that proposed by Hunt and Crossley \cite{Hunt.1975}, which rely on empirical parameters.
In the general case, therefore, three-dimensional finite element (FE) modeling has to be used.
A fine spatial resolution is needed to properly model the contact mechanics and the elastodynamics within the host structure.
Moreover, a fine time discretization is needed to properly resolve the collision events and subsequent wave propagation.
Periodic vibrations are the exception in the presence of impact events, so that many computationally efficient methods such as Harmonic Balance \cite{Krack.2019} are not applicable, and numerical time step integration has to be used instead.
Many vibration periods of the critical low-frequency modes need to be simulated in order to assess the steady-state vibration behavior for given initial conditions \cite{Theurich.2019,Li.2016b}.
Moreover, many of such simulations are generally needed to design \IESs, due to the nonlinear dependence on excitation conditions and design parameters.
With conventional FE Analysis (FEA) tools, this leads to an enormous computational effort.
\\
The computational effort of conventional FEA is driven by two aspects, the high mathematical model order (fine spatial discretization) and the artificial contact oscillations.
The latter are well-known to be due to the finite mass associated to every node at the contact interface using standard finite elements.
The velocity jumps occurring for contact activation and release then inevitably lead to high-frequency oscillations.
These are artificial in the sense that they do not occur in the spatially continuous case \cite{Khenous.2008}.
Therefore, strategies have been proposed for mass redistribution so that no inertia is associated to the nodal displacement degrees of freedom at the contact interface \cite{Khenous.2008,Hager.2008,Hager.2009,Renard.2010,Tkachuk.2013}.
This reduces the index of the differential-algebraic equation system from 3 to 1, and the momentum balance restricted to the contact boundary becomes a quasi-static sub-problem \cite{Acary.2008}.
Another important benefit of the massless-boundary approach is that no (phenomenological) impact law is needed for the boundary nodes.
The computational method recently developed by some of the authors of the present work combines the massless-boundary approach with component mode synthesis, in order to reduce both the spurious contact oscillations and the model order \cite{Monjaraz.2021}.
An important aim of the present study is to experimentally validate this computational method and assess its performance in comparison to conventional FEA.
The massless-boundary component mode synthesis method is applicable to both frictionless and frictional contact among multiple flexible bodies, but it is limited to linear elasticity and linear kinematics in its present form \cite{Monjaraz.2021}.
In spite of its broad applicability, we focus on the problem setting of a sphere impacting on a beam, motivated by the relevance of this scenario for impact vibration absorbers.

\subsection*{Experimental analysis of beams impacted by a sphere}
Many studies address the transverse impact of a sphere on a beam experimentally and confront the results with simulation \cite{Goldsmith.2001,Park.1994,schi2004,Dorogoy.2008,Qi.2016,Qi.2017,Wang.2017,Dong.2018,Zhang.2018}.
Good agreement was obtained \eg in \cite{schi2004,Seifried.2010,Qi.2017,Zhang.2018} in terms of displacement response and contact duration.
Modeling approaches range from three-dimensional FEA (\eg \cite{Qi.2017,Zhang.2018}) to modal superposition (\eg \cite{Park.1994}).
Seifried \etal \cite{schi2004} were one of the first to propose to combine modal superposition with local FE modeling of the contact region for impact simulations.
Contact was described using dynamic or quasi-static FE models, the Hertzian model or phenomenological models.
The focus of many studies was placed on energy dissipation by plastic deformation, whereas elastodynamic wave effects are neglected \cite{Stoianovici.1996,Qi.2016,Wang.2017,Dong.2018}, in complete contrast to the focus of the present study.
Moreover, the context of most studies was rigid multibody dynamics.
Consequently, a focus was placed on the macro-mechanical kinetic energy loss, commonly described by a coefficient of restitution.
In contrast, results on the modal energy distribution are not available in the literature for impacts of spheres on beams.
Similarly, experimental results on the contact force pulse (force vs. time) are lacking for this problem setting.
An important reason is that the contact force cannot be measured directly, and it is difficult to estimate it in the case of transversal impacts on beams, as opposed to the case of axial impacts on rods, for instance, where the force can be estimated from strain or velocity measurements at the free end, see \eg \cite{Cunningham.1985,Seifried.2010}.
Finally, results on the repeatability of the tests are only rarely reported.
In some studies, it was noted that sub-impacts lead to non-repeatability \cite{Stoianovici.1996,Seifried.2010,Qi.2016,Qi.2017}.
Sub-impacts denote multiple liftoff-collision events before the bodies rebound from each other for a longer period of time.
All these reasons motivated us to obtain our own experimental data for the validation.

\subsection*{Outline of present work}
The design of the test rig, the instrumentation and the test procedure are described in \sref{experiment}.
The computational methodology, including FE modeling, construction of the massless-boundary reduced-order model, and methods for numerical time integration and contact treatment are specified in \sref{simulation}.
The experimental results are confronted with computational predictions in \sref{evaluation}.
Conclusions are drawn in \sref{conclusions}.

\section{Experimental methodology}\label{sec:experiment}
The main aspects that contrast the presented experimental methodology from previous ones are the estimation of the contact force pulse and the modal energy distribution, as well as the measures taken to reduce and assess the repeatability.
First, we describe the test rig in its two configurations, free-free and clamped-clamped boundary conditions of the beam (\sref{design}).
We then describe the instrumentation and how the tests were carried out (\sref{procedure}).

\subsection{Design of the test rig}\label{sec:design}
To ensure a sound experimental database for the validation of the computational predictions, we defined the following requirements:
\begin{itemize}
\item[Req.1] high repeatability
\item[Req.2] insensitivity of macroscopic contact interactions to surface roughness and shape tolerances
\item[Req.3] relevance of wave propagation / multiple modes of vibration
\item[Req.4] well-separated natural frequencies
\item[Req.5] negligible plastic deformation
\item[Req.6] avoidance of sub-impacts
\item[Req.7] \MODIFIED{straight impact without rotation}
\end{itemize}
Plastic deformation is to be avoided both because of the technical relevance for \IESs and because of the aim for high repeatability.
Sub-impacts are to be avoided due to the otherwise high non-repeatability and uncertainty \cite{Stoianovici.1996,Seifried.2010,Qi.2016,Qi.2017}.
\\
Photos of the test rig are shown in \fref{BeamAndSphere}, a schematic illustration is given in \fref{setup}.
The setup consists of a flat and straight beam struck by a sphere.
Metal is selected as main construction material class; specifics and dimensions are listed in \tref{dimensions}.
Considering a slender beam ensures the relevance of wave propagation / multiple modes of vibration (Req.3).
The contact between a spherical and a flat interface is relatively insensitive to surface roughness and shape tolerances (Req.2).
A rectangular cross section of the beam is selected, which makes it easy to achieve well-separated natural frequencies (Req.4), in particular with regard to the bending modes in the two orthogonal directions.
The beam is manufactured of bright drawn steel, which is a well-known and well-accessible material, so that the experiment can be easily reproduced by other research groups.
The sphere is made of ferromagnetic, hardened steel and is taken from a ball bearing.
The theory in \cite{Christoforou.1998} is used to avoid sub-impacts by properly selecting the initial height and the geometric location of the impact for the given material combination, geometry and boundary conditions (Req.6).
\fig[t!]{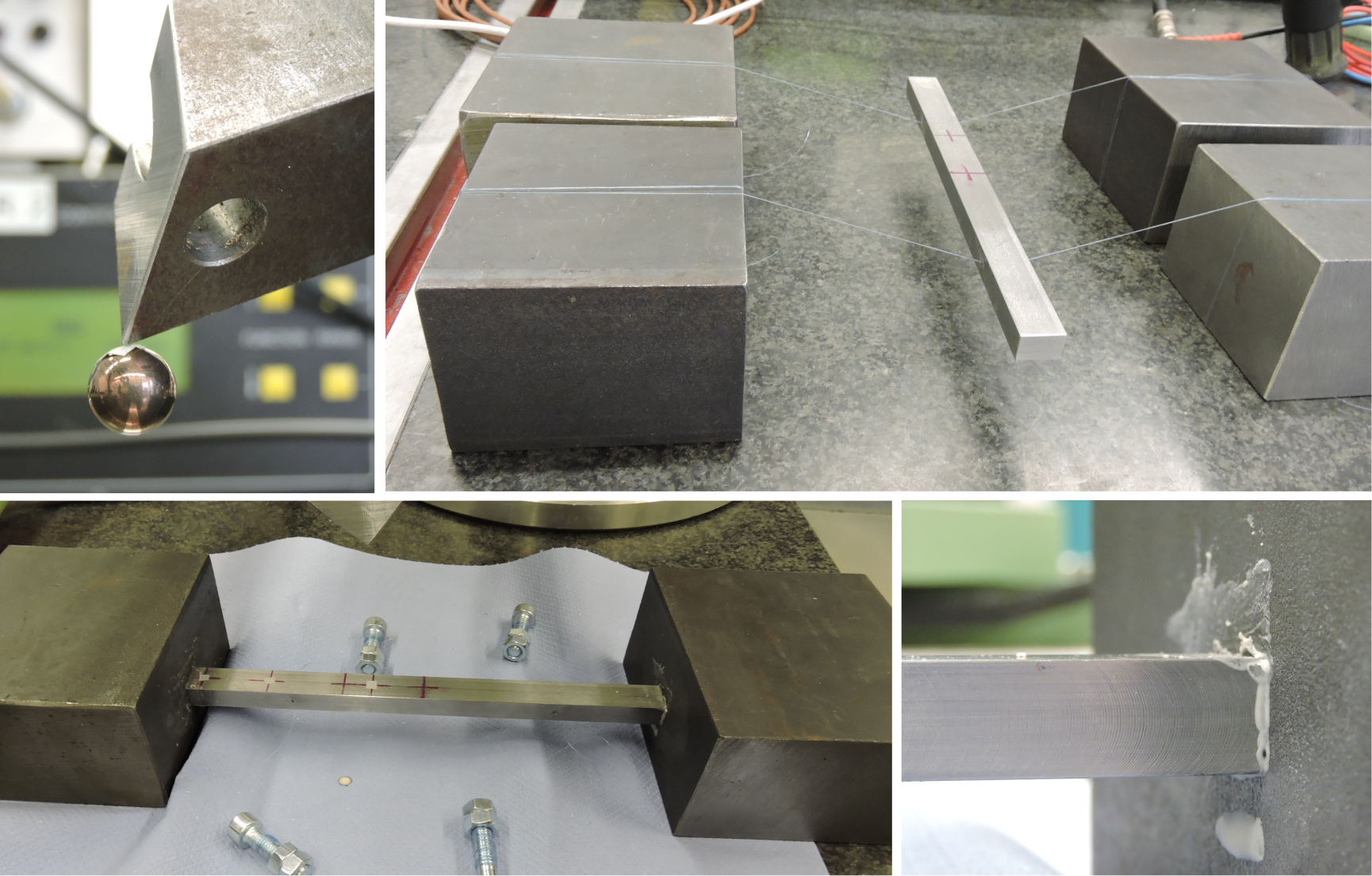}{Top left: Sphere suspended by magnet. Top right: Beam suspended by wires (free-free configuration). Bottom left: Beam glued to blocks (clamped-clamped configuration). Bottom right: Detail of gluing interface.
}{1.0}
\begin{table}[h]
	\centering
	\caption{Dimensions and material of beam and sphere selected for the test rig.}
	\begin{tabular}{|c||c|c|c|c|c|}
		\hline
		Beam & Length in mm & Width in mm & Height in mm & Weight in g & Material\\
		\hline
		& 210 & 15.0 & 10.0 & 247 & S235 (1.0038)\\
		\hline \hline
		Sphere & \multicolumn{3}{c|}{Diameter in mm} & Weight in g & Material\\
		\hline
		& \multicolumn{3}{c|}{11.1} & 5.58 & 100Cr6 (1.3505) \\
		\hline
	\end{tabular}
	\label{tab:dimensions}
\end{table}
\\
The desired impact velocities are achieved by dropping the sphere from a certain height.
The sphere is suspended by an electromagnet consisting of a coil and an iron core (\fref{setup}).
To ensure that the sphere hits the same contact point with the same velocity in a repeatable way (Req. 1), the iron core has a pointy shape (\fref{BeamAndSphere}-top-left).
Hence, the sphere's center of gravity and the pointy end of the iron core are exactly aligned in the direction of gravity.
\MODIFIED{
This ensures a repeatable impact point on the beam.
Moreover, this ensures a straight impact and avoids potential rotation (Req. 7) of the sphere.
This is regarded as important since the tangential motion and the sphere's rotation would be difficult to predict and measure.
}
\\
Two sets of boundary conditions are considered for the beam, free-free and clamped-clamped.
The boundary conditions lead to quite different wave propagation and reflection properties, which is regarded as important for a thorough validation of the simulations.
To mimic clamped-clamped boundary conditions, the beam is glued to massive steel blocks using an epoxy adhesive (\fref{BeamAndSphere}-bottom).
To mimic free-free boundary conditions, the beam is suspended by cords made of fishing wire (\fref{BeamAndSphere}-top-right), which is a rather common approach.
\MODIFIED{
The wires are made of polyamid and have a stiffness of about \ze{3000}{MPa}, which is two orders of magnitude lower than that of the beam.
}
The cords are passed through and then glued into \ze{1}{mm} drill holes to minimize the effect of friction.
The holes are placed at the vibration nodes of the beam's lowest-frequency bending mode in the horizontal symmetry plane.
Since this mode is expected to dominate the response, this further minimizes the cords' effect.
\\
\MODIFIED{
To further minimize the effect of the suspension by cords, especially on the beam's rigid body modes, one could consider to arrange the suspension in the vertical direction and analyze a horizontal impact.
Of course, the aforementioned strategy to obtain the desired impact velocity relies on gravity and therefore would have to be replaced, \eg by suspending the sphere via a long string as a pendulum \cite{Seifried.2010}.
In that case, however, the sphere must be modified (to attach the string), the trajectory of the sphere is no longer straight, and it seems practically impossible to ensure that the sphere is released in such a way that it hits the same contact point in a repeatable way (Req. 1) without rotation (Req. 7) and without introducing oscillations in the string direction (the string would have to be long and hence soft).
Because of these important drawbacks, this strategy (vertical suspension with horizontal impacts) was not further considered.
The results show very good agreement of the predictions obtained under ideal free-free boundary conditions with the test results (obtained under the described imperfections).
This justifies the selected suspension strategy a posteriori.
}
\\
The technical realizations of both the free-free and the clamped-clamped boundary conditions inevitably introduce damping to the system, which is very difficult to predict.
The damping is experimentally quantified in \sref{evaluation}.
As shown later, the resulting damping is negligible in the relatively short time span of interest immediately around the impact.

\subsection{Test procedure and instrumentation}\label{sec:procedure}
The considered standard steel shows plastic yield already for relatively low stresses. 
Thus, the contact area is plastically compressed and a shallow crater is formed.
Under repeated impacts, however, hardening occurs and a steady state is reached quickly, beyond which the plastic strain does not further increase significantly, and the contact region shows purely elastic behavior \cite{Seifried.2005} (Req. 5).
In the present study, this steady state was reached after at most 10 consecutive drops.
After this, the test was repeated another 10-11 times in order to analyze the repeatability.
\\
Two representative impact scenarios are analyzed in detail.
The scenarios differ in terms of boundary conditions (free-free vs. clamped-clamped beam), impact velocities, and impact locations, as specified in \tref{configs}.
The impact points are centered in the width direction but located at different axial positions.
The case of centric impact ($P_4$) is considered interesting, as it is expected to excite only odd bending modes.
In the case of eccentric impact, the impact point, $P_2$, is located so that it does not coincide with any of the nodes of the 7 lowest-frequency bending modes.
\begin{table}[h]
	\centering
	\caption{Tested impact scenarios. For definition of $h\si{d}$ and position $P_2$, $P_4$ see \fref{setup}.}
	\begin{tabular}{|c|c|c|c|}
		\hline
		suspension & impact point & dropping height $h\si{d}$ in mm & resulting impact velocity in mm/s\\
		\hline \hline
		free-free & $P_4$ (centric) & 73 & 1100 \\
		\hline
		clamped-clamped & $P_2$ (eccentric) & 42 & 769\\
		\hline
	\end{tabular}
	\label{tab:configs}
\end{table}
\begin{figure}[b!]
	\centering
	\includeinkscape[width=0.7\textwidth]{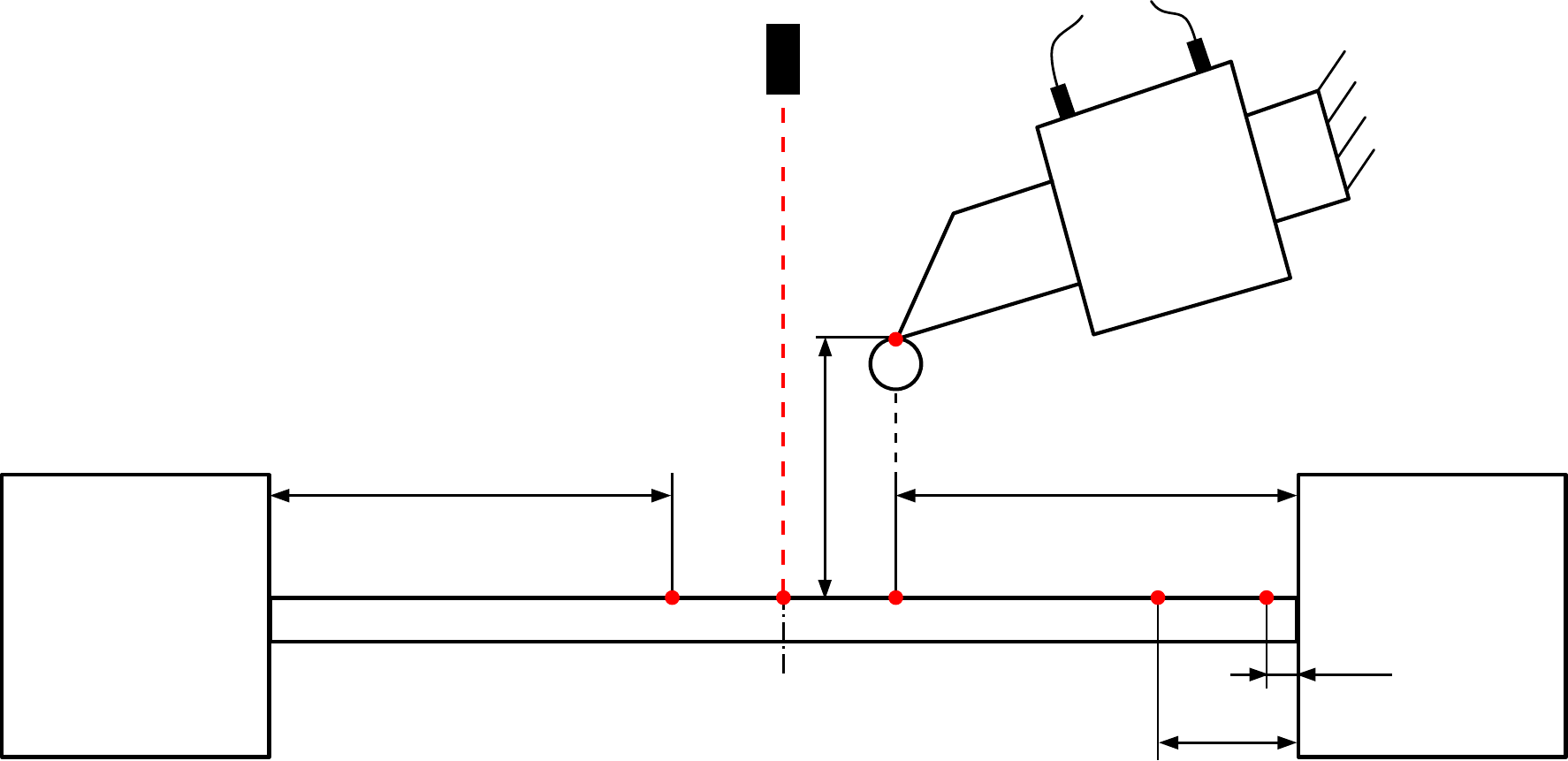} 
	\caption{Schematic illustration of the experimental setup in the clamped-clamped configuration and eccentric impact of the sphere.}
	\label{fig:setup}
\end{figure}
\\
The velocity response of beam and sphere are measured using laser-Doppler vibrometry (\fref{setup}).
Compared to other sensors, this permits a non-intrusive and highly accurate response measurement.
As shown later, the relevant frequency band goes up to about \ze{35}{kHz}.
The sampling rate was set to \ze{102.4}{kHz}, the highest value provided by the equipment.
The velocity component in the vertical direction is measured (\fref{setup}).
The measurement points are centered in width direction and equipped with reflection tape to improve the signal-to-noise ratio.
\\
To assess the quality of the underlying linear model of the beam for both sets of boundary conditions, the frequency response was experimentally determined, in addition to the drop tests.
To this end, impulse hammer testing was used.
The force applied at point $P_2^*$ (symmetric to $P_2$) is used as input, the response at point $P_2$ is used as output of the frequency response function.
To reduce the effect of noise, the test was repeated $N\si{rep}=10$ ($N\si{rep}=12$) times in the free-free (clamped-clamped) case, and the $H_1$-estimate is used (see \eg \zo{Randall.1987}),
\e{
H_1(f) = \frac{\sum\limits_{m = 1}^{N\si{rep}} U_m(f)\, F_{\mrm{exc},m}^* (f) }{\sum\limits_{m = 1}^{N\si{rep}} F_{\mrm{exc},m} (f)\, F_{\mrm{exc},m}^* (f)} \fk
}{H1}
where $f$ is the frequency, $U_m$ and $F_{\mrm{exc},m}$ denote the discrete Fourier transform of the $m$-th realization of the vertical displacement at point $P_2$ and the applied force, respectively, and $\Box^*$ denotes complex conjugate.
The Fourier transform of the displacement is obtained by integrating the Fourier transform of the measured velocity in the frequency domain (division by $2\pi f \ii$ with the imaginary unit $\ii=\sqrt{-1}$).
The $H_1$-estimate is particularly suitable if the response is more noisy than the excitation \zo{Randall.1987}, which is the case for the present measurements.

\section{Computational methodology}\label{sec:simulation}
The point of departure for the computational predictions is the geometry of the test rig in its two configurations.
The assumptions on boundary and initial conditions, as well as contact and material properties are described in \sref{assumptions}.
A three-dimensional FE model is then constructed (\sref{mesh}).
Three different simulation approaches were pursued and assessed.
The first and second approaches are implicit and explicit time step integration of the FE model, respectively, carried out using the commercial tool \ABAQUS (state-of-the-art).
The third approach is a recently developed technique \cite{Monjaraz.2021} implemented in \MATLAB.
{
Its main innovative aspect is to achieve a massless contact boundary using component mode synthesis, which reduces the mathematical model order and numerical oscillations (compared with conventional FEA), leading to excellent energy conservation and computational performance.
}

\subsection{General assumptions: boundary and initial conditions, contact and material properties}\label{sec:assumptions}
The FE model of the sphere and the impacted structure are depicted for the clamped-clamped configuration in \fref{model}.
The impacted structure not only consists of the beam, but it also includes the blocks and the adhesive connecting the beam with the blocks.
The bottom surface of the blocks is assumed to be tied to the ground.
In the free-free configuration, the fishing wires and the small holes through which the wires are passed and glued are not modeled, so that the impacted structure is only the beam.
\begin{table}[b]
	\centering
	\caption{Material properties}
	\begin{tabular}{|c||c|c|c|}
		\hline
		Material & Elastic modulus $E$ in MPa & Poisson's ratio $\nu$ & density $\rho$ in $\mrm{kg/m^3}$ \\
		\hline \hline
		Steel & $210,000$ & $0.3$ & $7,800$ \\
		\hline
		Adhesive & $7,000$ & $0.3$ & $950$\\
		\hline
	\end{tabular}
	\label{tab:materials}
\end{table}
\begin{figure}[t!]
	\centering
	\includeinkscape[width=0.65\textwidth]{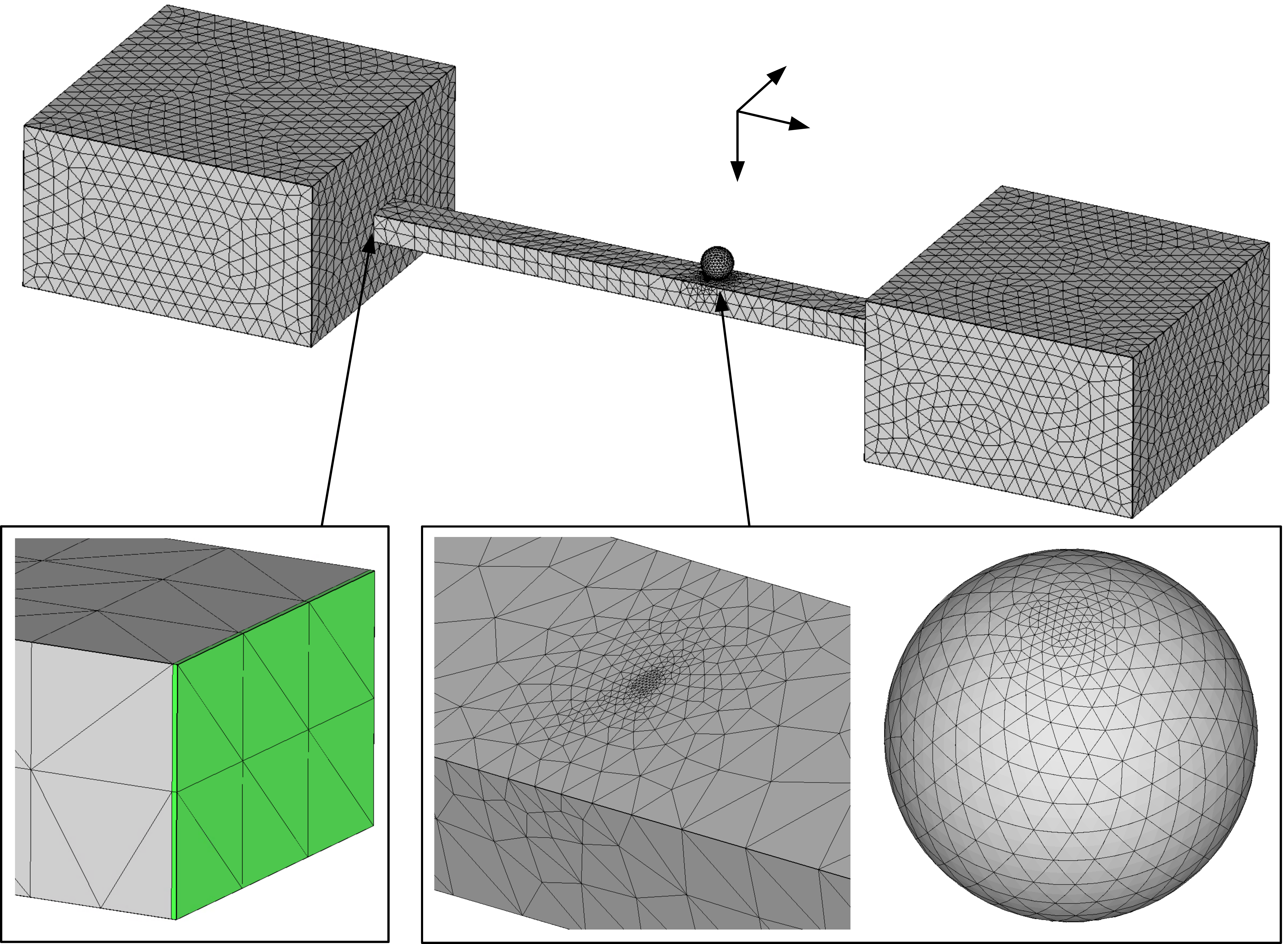}	
	\caption{FE model in the clamped-clamped configuration. Bottom left: The element row representing the adhesive is shown in green. Bottom right: Locally refined mesh of beam and sphere in contact region.}
	\label{fig:model}
\end{figure}
\\
Instead of simulating the falling of the sphere from its point of release, in the model, the sphere is placed close to the beam and given an according initial velocity (as listed in \tref{configs}).
It was found that damping and gravity forces are negligible in the considered relatively short time interval and therefore not accounted for in the simulation.
\\
Homogeneous, linear-elastic and isotropic material behavior is assumed for all solid bodies.
As described in \sref{experiment}, beam, sphere and blocks are made of steel, the properties of which are listed in \tref{materials}.
The properties of the \ze{0.2}{mm} thick adhesive layer (clamped-clamped configuration) is also listed in \tref{materials}.
While the properties of steel are well-known, there is no reliable data available for the adhesive.
Thus, the Young's modulus and the density of the adhesive were heuristically tuned to achieve reasonable agreement of model and experiment in terms of natural frequencies.
\\
Between sphere and beam, frictionless unilateral contact interactions are modeled.
It seems justified to neglect friction because the sphere release mechanism ensures that the sphere impacts without spin.
The aforementioned plastic crater is not modeled but instead the initial geometry is considered with nominal material properties.

\subsection{FE mesh generation}\label{sec:mesh}
The FE mesh is depicted in \fref{model}.
Sphere, beam and blocks are discretized using second-order tetrahedral elements (C3D10).
The adhesive layer is discretized using second-order prismatic elements (C3D15), which were extruded from the beam interfaces and tied to the blocks.
The mesh must be fine enough to properly resolve both contact interactions and propagating waves.
For wave propagation, a rule of thumb is to sample the shortest relevant wave length with at least 20 nodes \zo{Moser.1999}.
The minimum wave speed in a homogenous elastic medium is that of transversal waves,
{
$c\si{T}$, and it depends on the elastic modulus $E$, Poisson ratio $\nu$ and density $\rho$ as follows
}
\zo{Geradin.2014}:
\e{
c\si{T} = \sqrt{\frac{E}{2 (1+\nu) \rho}}\fp
}{waveSpeed}
The measurements indicate that the highest relevant frequency is about $f\si{max} = \ze{35}{kHz}$ (\cf convergence study described in \ssref{ROMsimulation} and experimentally obtained modal energy distribution shown in \sssref{modalEnergyDistribution} ).
This leads to a minimum wave length of $c\si{T}/f\si{max}=92~\mrm{mm}$ for steel.
Taking into account that the (second-order) elements used in this work have mid-side nodes, the element length should not exceed $9.2~\mathrm{mm}$.
We set the nominal element length to $\ze{6}{mm}$.
To properly resolve the local flexibility in the contact region, the mesh was gradually refined towards the center of the contact area (\fref{model}-bottom-right).
Based on a mesh convergence study, a refinement down to an element length of $\ze{0.2}{mm}$ (beam) and $\ze{0.3}{mm}$ (sphere) was used for the results presented in this work.
The resulting model contains $42,651$ elements with a total of $65,321$ nodes in the clamped-clamped configuration, and $14,551$ elements with a total of $22,020$ nodes in the free-free one.
\\
In conjunction with explicit time step integration, \ABAQUS does not support C3D10 elements, so that we used their modified, lumped-mass variant (C3D10M).
Mass lumping leads to a diagonal mass matrix, which is essential for the efficiency of the explicit algorithm \zo{Abaqus.2016}.
Moreover, the C3D15 elements, used to model the adhesive in the clamped-clamped configuration, are incompatible with explicit time step integration.
As \ABAQUS does not provide a lumped-mass alternative to the C3D15 element, a complete rework of the base model would have been necessary to apply the explicit algorithm.
Because of these difficulties, the clamped-clamped configuration was only analyzed using the implicit algorithm in this work.
The free-free configuration was analyzed using both the explicit and the implicit algorithm.

\subsection{Simulation using the FE model}\label{sec:FEsimulation}
Details on the implicit and explicit time step integration methods, and the contact treatment are described in the following.

\subsubsection{Implicit integration of FE model and contact treatment}
For implicit time step integration, the Hilber-Hughes-Taylor scheme \zo{Hilber.1977} was used with \ABAQUS default parameters for contact problems \zo{Abaqus.2016}. 
As a rule of thumb, the shortest relevant oscillation period should be sampled with at least $20$ time steps.
With the aforementioned highest relevant frequency, $f\si{max} = \ze{35}{kHz}$, this leads to a maximum step size of $\Delta t = \ze{1.43\cdot 10^{-6}}{s}$.
A smaller maximum step size of $\Delta t = \ze{1\cdot 10^{-7}}{s}$ was selected, as it was found that the impact event is otherwise not properly resolved.
\\
A segment-to-segment contact formulation is used, in conjunction with an augmented Lagrangian scheme using default parameters for the relative penetration tolerance and the number of allowed augmentations \zo{Abaqus.2016}. 
Tightening the relative penetration tolerance did not yield significantly different results, but only led to convergence problems.

\subsubsection{Explicit integration of FE model and contact treatment}
For explicit time step integration, the central difference scheme was used.
The time step was automatically selected to achieve numerical stability (Courant-Friedrichs-Lewy condition) \zo{Abaqus.2016}. 
This led to an average time step size of $\Delta t = \ze{6.58\cdot 10^{-9}}{s}$, which is more than one order of magnitude smaller than the maximum time step size used for implicit integration.
\\
A node-to-segment contact formulation with balanced independent-dependent algorithm was used. 
To kinematically enforce the contact constraints, the Lagrange multiplier technique was used with a predictor-corrector scheme.

\subsection{Simulation using massless-boundary reduced-order model (ROM)}\label{sec:ROMsimulation}
The simulation using massless-boundary ROM relies on MacNeal's component mode synthesis method and a semi-explicit time step integration scheme.
{
Due to the massless boundary, the contact problem simplifies to a quasi-static one.
The static contact problem is solved implicity using an augmented Lagrangian scheme, while the dynamic sub-problem is marched explicitly for efficiency.
}

\subsubsection{ROM construction using MacNeal method}
We propose to achieve the massless contact boundary by appropriate component mode synthesis \cite{Monjaraz.2021}.
This permits to use standard finite elements for the parent model, in contrast to previous works which propose to modify the location of the integration points \cite{Hager.2008,Hager.2009} or adjust the shape functions \cite{Renard.2010,Tkachuk.2013}.
Another important advantage is that component mode synthesis reduces the model order compared to the parent FE model.
\\
For component mode synthesis, the MacNeal method is used in this work \cite{macn1971}.
First, the model is split into nodal degrees of freedom associated with the contact (boundary), $\qqb$, and the remaining (inner) degrees of freedom, $\qqi$.
The nodal degrees of freedom are approximated as a linear combination of a set of component modes, 
\e{
	\qq = \vector{\qqb\\ \qqi} \simeq  \mm R \vector{\qqb \\ \mm\eta} \fp
}{MacNeal}
The component modes are the columns of the matrix $\mm R$.
In the MacNeal method, the component modes are a subset of free-interface normal modes and residual flexibility attachment modes \cite{macn1971}.
Note that the boundary coordinates, $\qqb$, are retained, while the inner coordinates, $\qqi$, are replaced by the coordinates, $\mm\eta$, of the retained free-interface normal modes (where usually $\dim \mm\eta \ll \dim\mm\qqi$).
The component modes are determined from the mass matrix, $\mm M$, and the stiffness matrix, $\mm K$, of the FE model.
We exported the FE matrices from \ABAQUS and carried out the model order reduction in \MATLAB.
The MacNeal method yields reduced mass and stiffness matrices, which are defined, along with $\mm R$ in \aref{MacNeal}.
An important property of the MacNeal method is that the static flexibility with respect to loads applied at the boundary is described as accurately as the parent FE model \cite{macn1971}.
To properly model the elastodynamics, the retained normal modes should cover the relevant frequency range of the response.
The MacNeal method readily yields a singular mass matrix, where no inertia is associated with the boundary coordinates.
\\
The MacNeal method is applied, individually, to the sphere and to the impacted structure. 
The dimensions of the contact area were estimated using Hertzian theory and a nominal contact area was specified that is sufficiently large to contain all active contact nodes.
The sphere has 48 (59) and the beam 41 (42) nodes in the nominal contact area in the free-free (clamped-clamped) configuration.
The difference between both configurations stems from the automatic mesh generation.
{
Based on the results of a convergence study,
}
all normal modes with natural frequency in the range from \ze{0}{Hz} to \ze{69.7}{kHz} (\ze{40.8}{kHz}) are retained in the free-free (clamped-clamped) configuration.
With this, the ROM of the beam contains 6 rigid body modes and 40 elastic modes in the free-free configuration, and 80 elastic modes in the clamped-clamped configuration.
Note that the clamping blocks lead to a higher modal density.
The ROM of the sphere contains only the 6 rigid body modes since its elastic modes start from $\ze{230}{kHz}$.
As a consequence, the deformation of the sphere is modeled in a quasi-static way.

\subsubsection{Semi-explicit integration and contact treatment}
For time step integration, we used the Verlet scheme, which is second-order accurate (fixed contact conditions), time-reversible and has favorable energy conservation properties (symplectic integrator).
We use the notation $\qqb^n = \qqb(t^n)$, where $t^n$ is the $n$-th time level.
Applied to the massless-boundary ROM subjected to frictionless contact, we obtain the constrained equation system \cite{Monjaraz.2021}
\ea{
\kbb\qqb^n + \kbi\mm\eta^n - \mm W\mm \lambda^n &=& \mm 0 \fk \label{eq:algebraic} \\
\frac{\dot{\mm\eta}^{n+\frac12} - \dot{\mm \eta}^{n-\frac12}}{\Delta t} + \kbitra\qqb^n + \kii\mm\eta^n &=& \mm 0 \fk \label{eq:differential} \\
\mm\eta^{n+1} &=& \mm\eta^n + \dot{\mm\eta}^{n+\frac12}\Delta t \fk \label{eq:explicit} \\
\mm 0 \leq \mm g^n = \mm W\tra\qqb^n & \perp & \mm\lambda^n \geq \mm 0\fp \label{eq:complementary}
}
Herein, $\kbb$, $\kbi$ and $\kii$ are sub-matrices of the reduced stiffness matrix, $\Ktil$, defined in \erefs{MacNealKbb}-\erefo{MacNealKii}.
$\mm\lambda$ is the vector of Lagrange multipliers which can be interpreted as contact forces, with the constant matrix of contact force directions $\mm W$.
Frictionless unilateral contact conditions are expressed as complementary inequality constraints in \eref{complementary}, without introducing any empirical parameters (\eg contact stiffness).
Herein, $\perp$ means that if a certain element of the contact gap vector $\mm g^n$ is nonzero, then the corresponding element of $\mm\lambda^n$ must be zero, and vice versa.
The algorithm is as follows:
For given $\mm\eta^n$, \eref{algebraic} is solved for $\qqb^n$ subjected to the constraints in \eref{complementary}, using an augmented Lagrangian technique in conjunction with Jacobi relaxation \cite{Monjaraz.2021}. 
Subsequently, \eref{differential} is solved explicitly for $\dot{\mm\eta}^{n+\frac12}$, followed by the update in \eref{explicit}.
Then, $n$ is increased by one until the end of the simulation is reached.
Since the sub-problem restricted to the boundary coordinates (static contact problem) is solved implicitly, and the sub-problem restricted to the inner coordinates (dynamic internal force balance) is explicitly integrated, we refer to the algorithm as semi-explicit \cite{Monjaraz.2021}.
Details on the implementation of the algorithm are given in \cite{Monjaraz.2021}.
The time step is set as in the case of the implicit FEA ($\Delta t=\ze{1\cdot 10^{-7}}{s}$) for fair comparison.
Numerical studies showed that the time step can be increased up to $\Delta t = \ze{2.5\cdot 10^{-7}}{s}$ without significantly changing the presented results.
\\
A node-to-segment contact formulation with constant node pairing is used, where the beam is treated as independent and the sphere as dependent.
As the implementation is currently limited to first-order contact elements, the mid-side nodes at the contact interface were kinematically constrained to their associated vertex nodes.
These constraints increase the stiffness of the contact interface slightly, and it is expected that the contact stress distribution is only poorly resolved in this way.
As shown in \sref{evaluation}, however, the structural dynamic response (on velocity level), and in particular the modal energy distribution is predicted with excellent accuracy.

\subsection{Opportunities and Limitations of Hertzian contact modeling in conjunction with modal truncation}\label{sec:HertzianPlusModal}
For the particular problem setting of a sphere impacting a beam, modal truncation in conjunction with a Hertzian contact model \cite{hert1881} seems well-suited \cite{Seifried.2010}.
This is especially true because an important design goal of \IESs is to achieve purely elastic material behavior in the contact region (important assumption of Hertzian theory).
However, Hertzian theory is limited to the frictionless case and to problem settings where the contact radius is small compared to the dimensions of the bodies (beam thickness and sphere radius in the considered problem setting).
In particular, conforming contact, such as flat on flat, cannot be properly modeled using Hertzian theory.
These are important limitations for the design of \IESs.
Consequently, this type of modeling approach is not further analyzed in the present work.

\section{Evaluation}\label{sec:evaluation}
In this section, measurements are confronted with predictions obtained using the FE model and the massless-boundary ROM.
For clarity in the figures, results of either the initial FE model (used for implicit time integration) or the lumped-mass variant (used for explicit time integration) are shown.
If not stated otherwise, results of the free-free configuration are depicted only for the lumped-mass variant, referred to as \emph{FE model (explicit)}.
Accordingly, results of the clamped-clamped configuration are depicted only for the initial FE model, denoted \emph{FE model (implicit)}.
First, the linear model of the impacted structure is analyzed.
Then, the impact process and the resulting dynamics are analyzed.
%
%

\subsection{Linear frequency response and modal analysis}\label{sec:linearFRF}
To characterize the linear behavior of the impacted structure, its frequency response function, $H(f)$, was determined between force input at point $P_2^*$ and displacement output at point $P_2$ (\cf labels in \fref{setup}).
As described in \sref{procedure}, $H(f)$ was experimentally determined using repeated impulse hammer testing along with the $H_1$-estimate defined in \eref{H1}.
Further, a conventional modal analysis was applied to extract natural frequencies, modal damping ratios and mass-normalized modal deflection shapes.
The magnitude of $H(f)$ is depicted in \fref{H1}.
{
As expected for lightly-damped, flexible structures, $\left| H(f) \right|$ contains several sharp resonance peaks and anti-resonances.
}
The extracted natural frequencies up to about $35~\mathrm{kHz}$ of the bending modes in the vertical direction (dropping direction; $y$-direction in \fref{model}) are listed along with the associated modal damping ratios in \tref{f_and_D}.
{
The bending modes in the vertical direction (flap-wise) are denoted as 1F through 9F (sorted by ascending natural frequency).
}
\\
Damping is extremely low in the free-free configuration, as it is mainly caused by material dissipation and the ambient air.
Damping is higher, but still very low in the clamped-clamped configuration.
Here, the higher damping is attributed to the viscoelastic behavior of the adhesive and dry friction between clamping blocks and ground.
Recall that the Young's modulus and the density of the adhesive were heuristically updated in the model to achieve a reasonable match of the experimentally obtained natural frequencies.
The experimentally obtained modal damping ratios were adopted in the model to generate \fref{H1}, to facilitate the comparison.
It should be emphasized that this is done only in \fref{H1}, whereas damping is neglected in the model for the subsequent analysis of the impact process and the resulting dynamics.
\begin{figure}[t!]
\centering
	\begin{minipage}{\textwidth}
		\includegraphics[width=\textwidth]{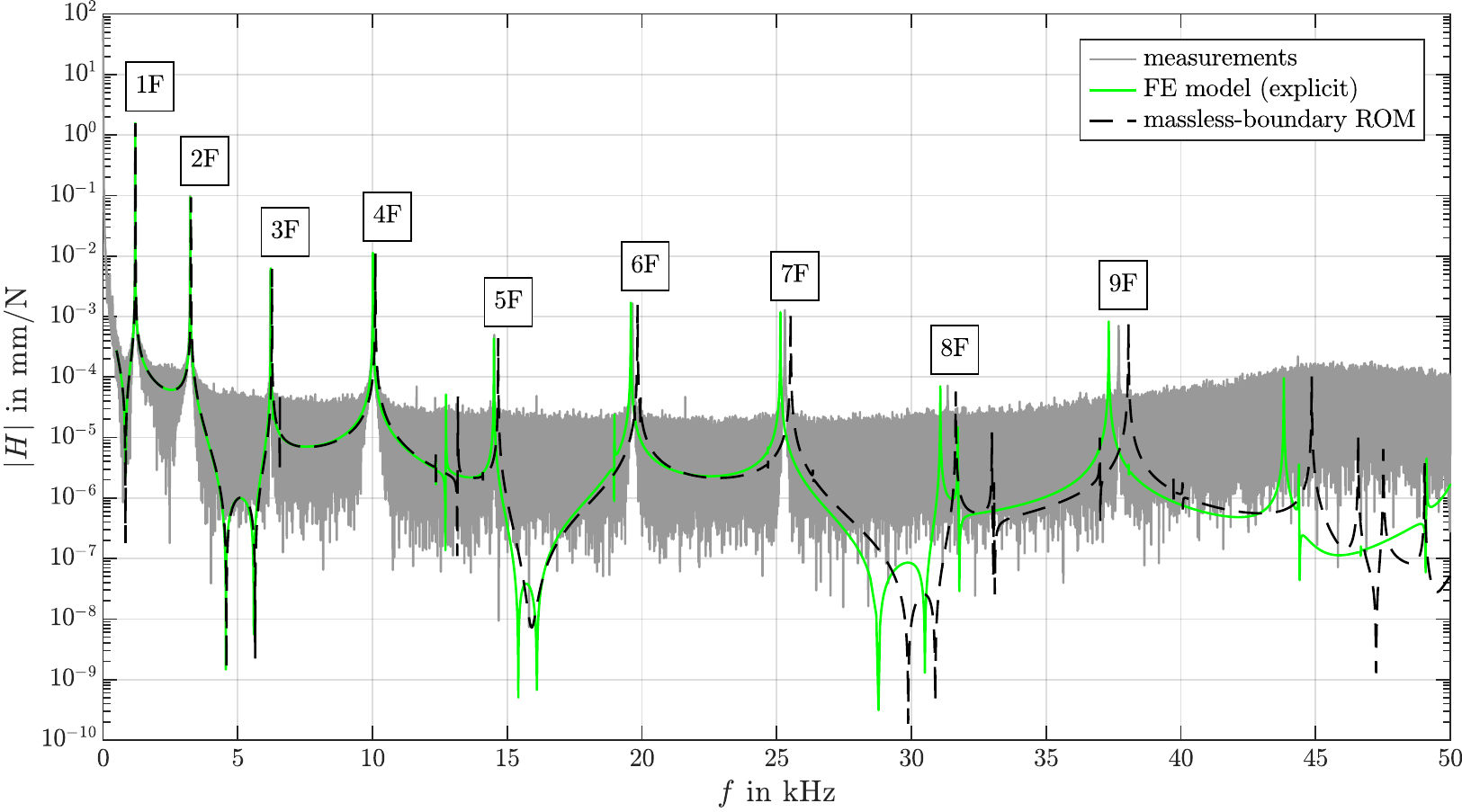}
	\end{minipage}\\
	\begin{minipage}{\textwidth}
		\includegraphics[width=\textwidth]{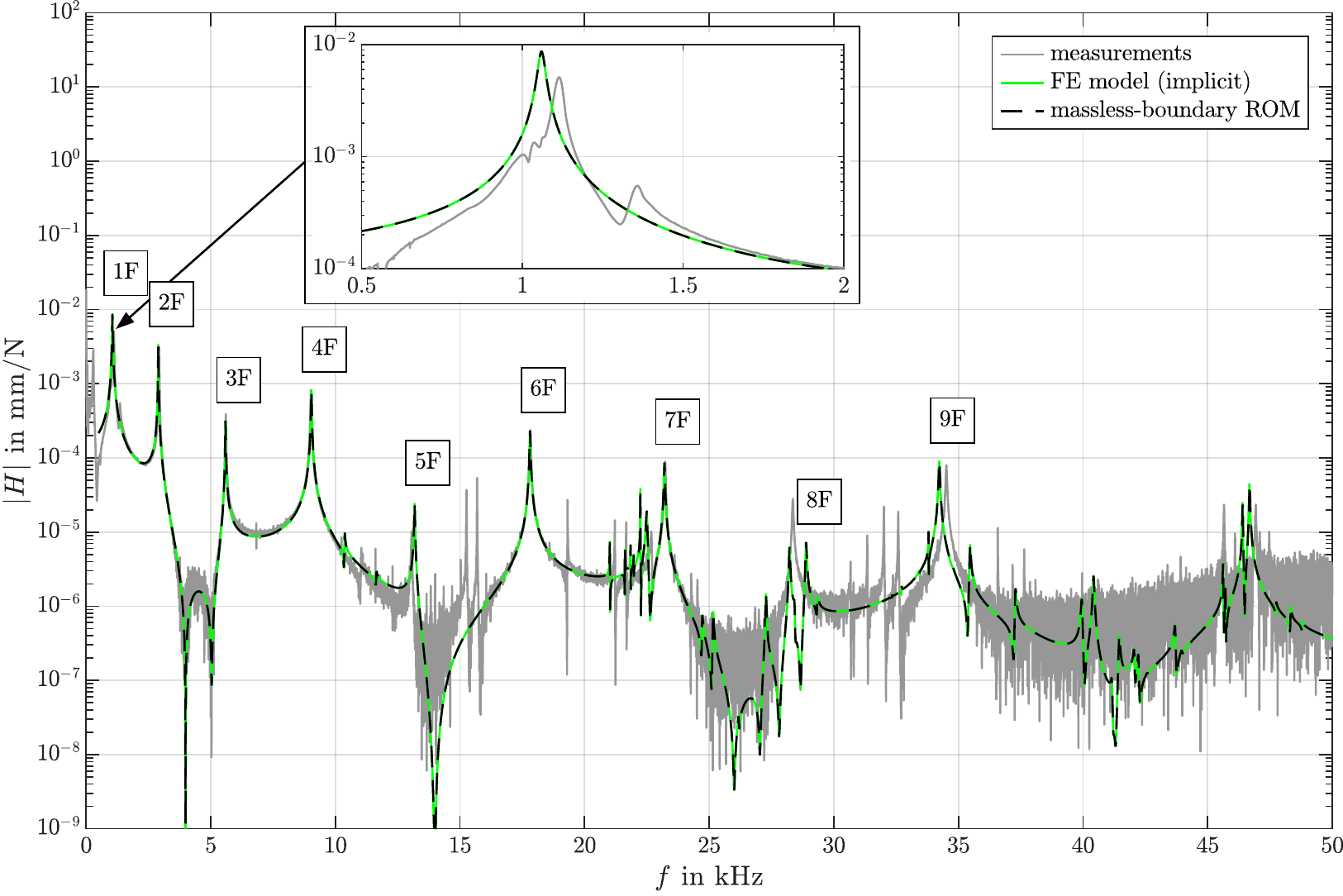}
	\end{minipage}
	\caption{Frequency response function obtained from measurement and simulation. Top: Free-free configuration. Bottom: Clamped-clamped configuration.}
\label{fig:H1}
\end{figure}
\begin{table}[h]
	\centering
	\caption{Natural frequencies $f_k = \omega_k/(2\pi)$ and damping ratios $D_k$ of the bending modes in vertical direction, extracted from the experimentally obtained frequency response function. The 8F mode is only poorly excited in the free-free configuration, which is why the entry is missing.}
	\begin{tabular}{|c||c|c|c|c|}
		\hline
		 & \multicolumn{2}{c|}{free-free} & \multicolumn{2}{c|}{clamped-clamped} \\
		\hline
		mode & $f$ in kHz & $D$ in $10^{-6}$ &  $f$ in kHz & $D$ in $10^{-3}$ \\
		\hline
		1F & 1.19 & 22.9 & 1.11 & 10.3 \\
		\hline
		2F & 3.23 & 53.2 & 2.93 & 1.93 \\
		\hline
		3F & 6.21 & 59.4 & 5.60 & 1.52 \\
		\hline
		4F & 10.0 & 85.0 & 9.02 & 1.32 \\
		\hline
		5F & 14.5 & 90.1 & 13.1 & 1.02 \\
		\hline
		6F & 19.6 & 89.8 & 17.8 & 0.830 \\
		\hline
		7F & 25.3 & 81.0 & 23.2 & 0.727 \\
		\hline
		8F & - & - & 28.4 & 0.736 \\
		\hline
		9F & 37.7 & 72.8 & 34.5 & 0.643 \\
		\hline
	\end{tabular}
	\label{tab:f_and_D}
\end{table}
\\
Almost all analyzed natural frequencies of the vertical bending modes, the overall frequency response function, along with the anti-resonances match very well between simulation and measurement.
Looking closely, some discrepancy can be encountered near the fundamental vertical bending mode (1F) in the clamped-clamped configuration.
Here, the estimated frequency response function has an \myquote{unclean} peak, leading to uncertainty in the identification of the modal properties.
As a consequence, the corresponding mass-normalized mode shape is by a factor of 1.19 higher in the model.
Moreover, the corresponding natural frequency deviates by $5\%$, while the relative error remains $\leq 1.2\%$ for all other analyzed vertical bending modes in both configurations.
\\
From the results of the free-free configuration (\fref{H1}-top), it is apparent that the mass lumping required for explicit FEA leads to slightly different, typically lower natural frequencies.
The results of the FE model without mass lumping (depicted only in clamped-clamped configuration) are practically indistinguishable from those of the massless-boundary ROM.
\\
Regarding the clamped-clamped configuration (\fref{H1}-bottom), there are several smaller resonance peaks between the bending-dominated modes.
These correspond to modes dominated by the clamping blocks.
Some of these are not well matched, which is attributed to the simplified modeling of the contact between the clamping blocks and the ground.
\\
Overall, it is concluded that the frequency response functions match very well between simulation and measurements, and thus the linear FE models and the ROM yield very good approximations in the relevant frequency range from $0~\mrm{kHz}$ to $35~\mrm{kHz}$.

\subsection{Analysis of the impact event and the resulting dynamics}\label{sec:impact}
In this section, the nonlinear structural dynamic behavior of the sphere and the impacted structure is investigated.
The analysis includes the velocity response of beam and sphere, the contact force pulse and the post-impact modal energy distribution.
Measurements are compared against computational predictions based on the massless-boundary ROM and FEA.
Finally, the speedup achieved by the massless-boundary ROM over the FEA is discussed.

\subsubsection{Velocity response}
\begin{figure}[t!]
	\begin{minipage}{\textwidth}
		\begin{minipage}{0.49\textwidth}
			a)\vfill
			\includegraphics[width=\textwidth]{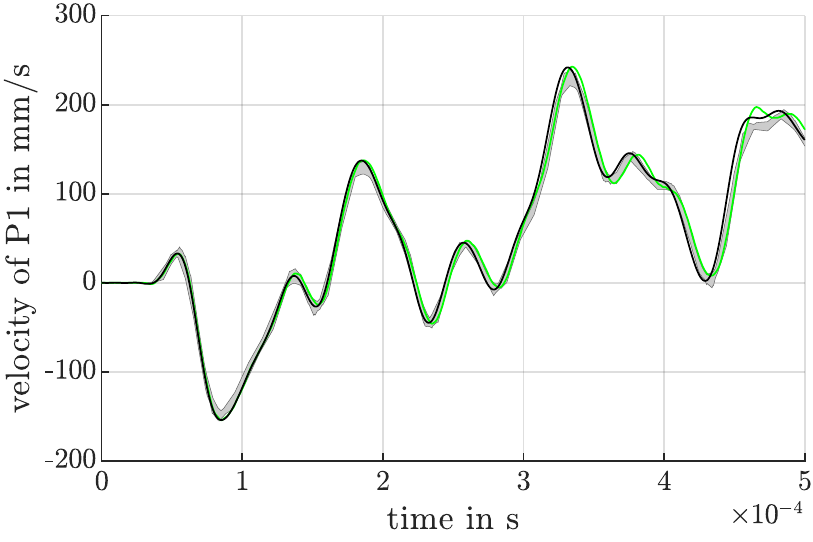}
		\end{minipage}
		\hfill
		\begin{minipage}{0.49\textwidth}
			b)\vfill
			\includegraphics[width=\textwidth]{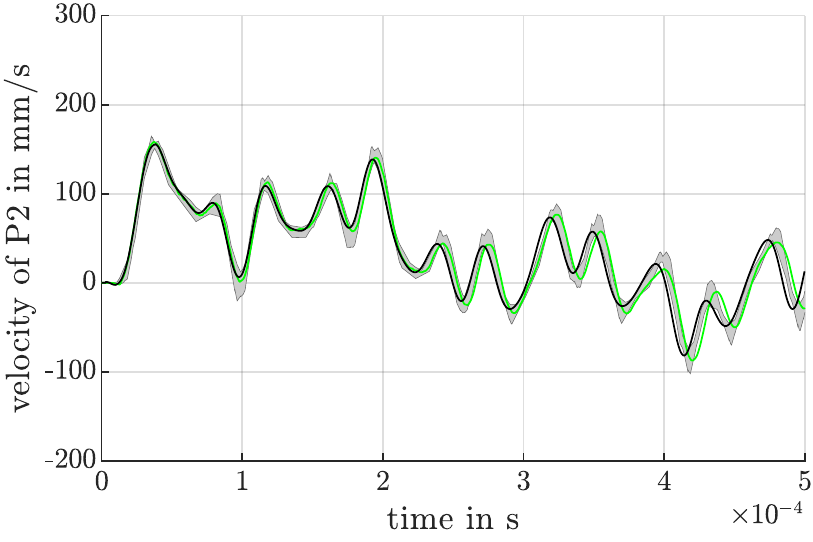}
		\end{minipage} \\
		\begin{minipage}{0.49\textwidth}
			c)\vfill
			\includegraphics[width=\textwidth]{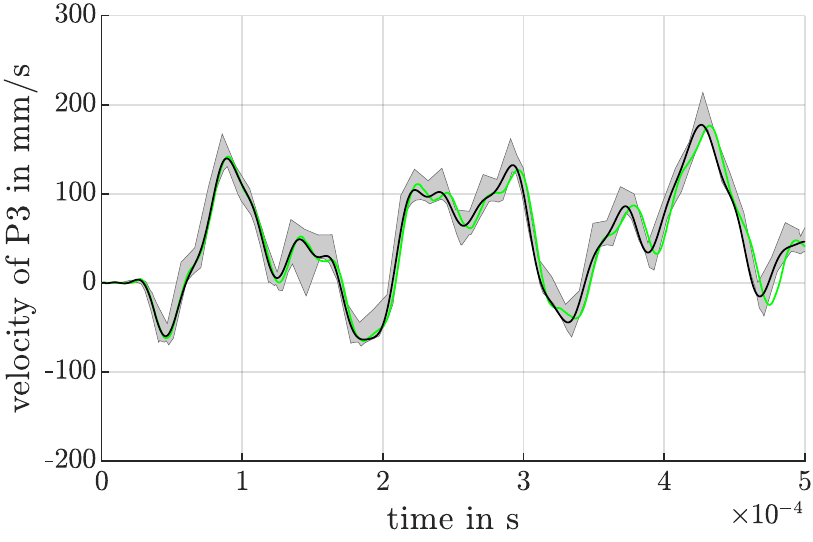}
		\end{minipage}
		\hfill
		\begin{minipage}{0.49\textwidth}
			d)\vfill
			\includegraphics[width=\textwidth]{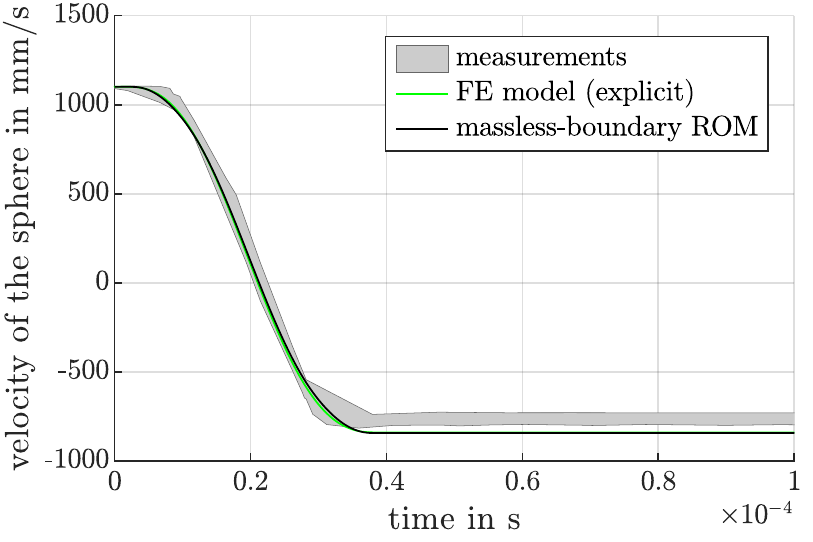}
		\end{minipage}
	\end{minipage}\\
	\begin{minipage}{\textwidth}
		\begin{minipage}{0.49\textwidth}
			e)\vfill\centering
			\includeinkscape[width=0.8\textwidth]{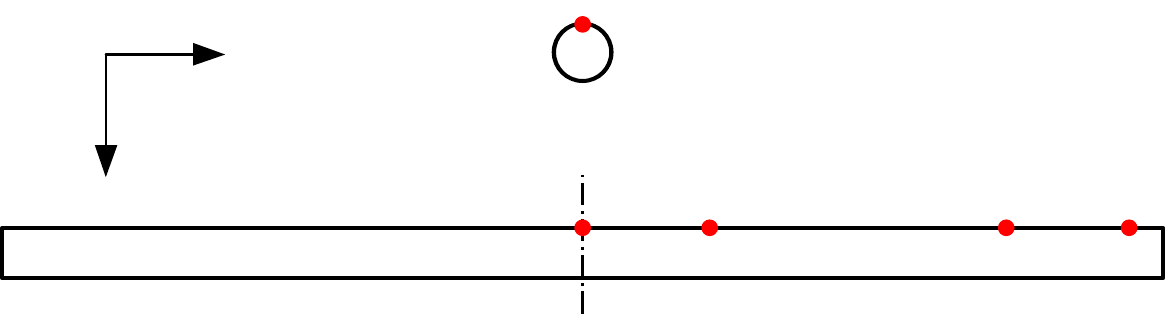}
		\end{minipage}
		\hfill
		\begin{minipage}{0.49\textwidth}
\caption{
Velocity response under central impact in the free-free configuration.
a) velocity of $P_1$,
b) velocity of $P_2$,
c) velocity of $P_3$,
d) velocity of $P\si{sph}$,
e) illustration of problem setting.
}
			\label{fig:val_free}
		\end{minipage}
	\end{minipage}
\end{figure}
\begin{figure}[h!]
	\begin{minipage}{0.49\textwidth}
		a)\vfill
		\includegraphics[width=\textwidth]{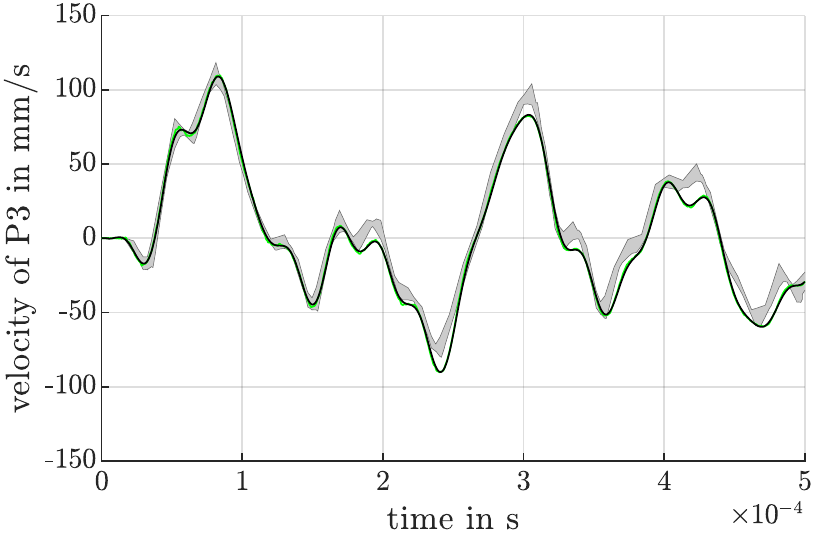}
	\end{minipage}
	\hfill
	\begin{minipage}{0.49\textwidth}
		b)\vfill
		\includegraphics[width=\textwidth]{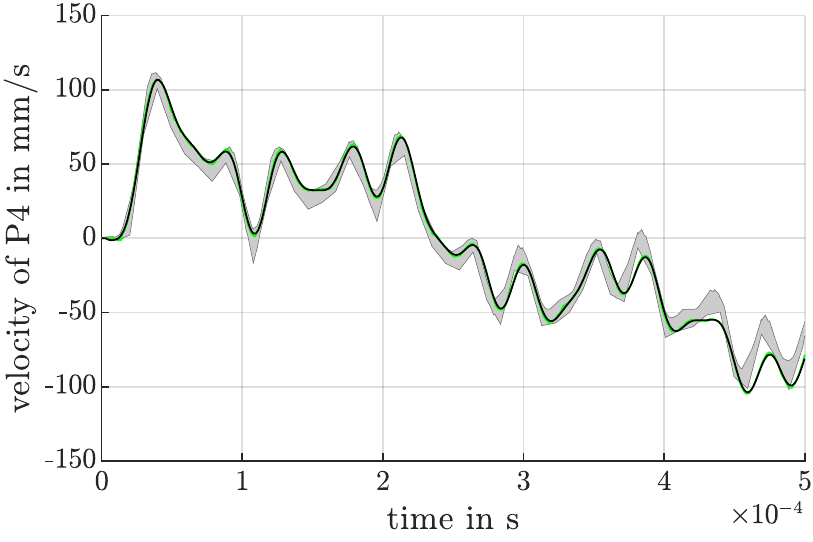}
	\end{minipage} \\
	\begin{minipage}{0.49\textwidth}
		c)\vfill
		\includegraphics[width=\textwidth]{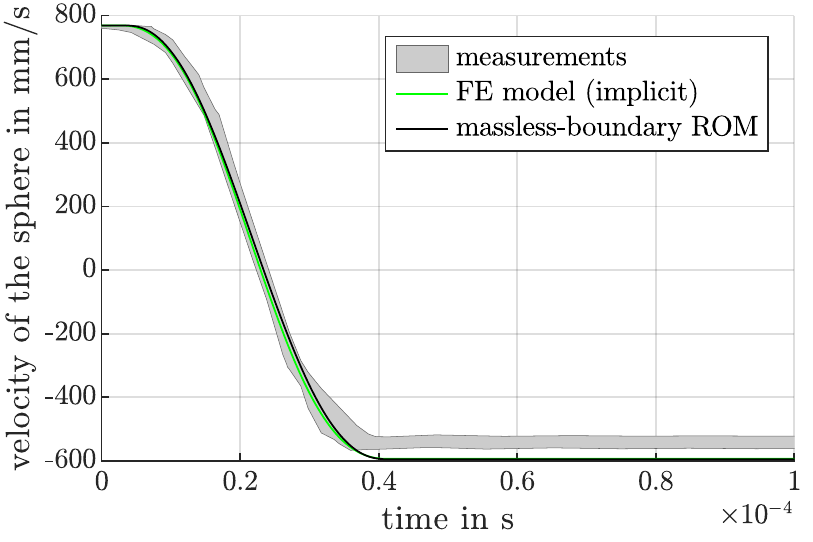}
	\end{minipage}
	\hfill
	\begin{minipage}{0.49\textwidth}
		d)\vfill
		\centering
		\includeinkscape[width=0.8\textwidth]{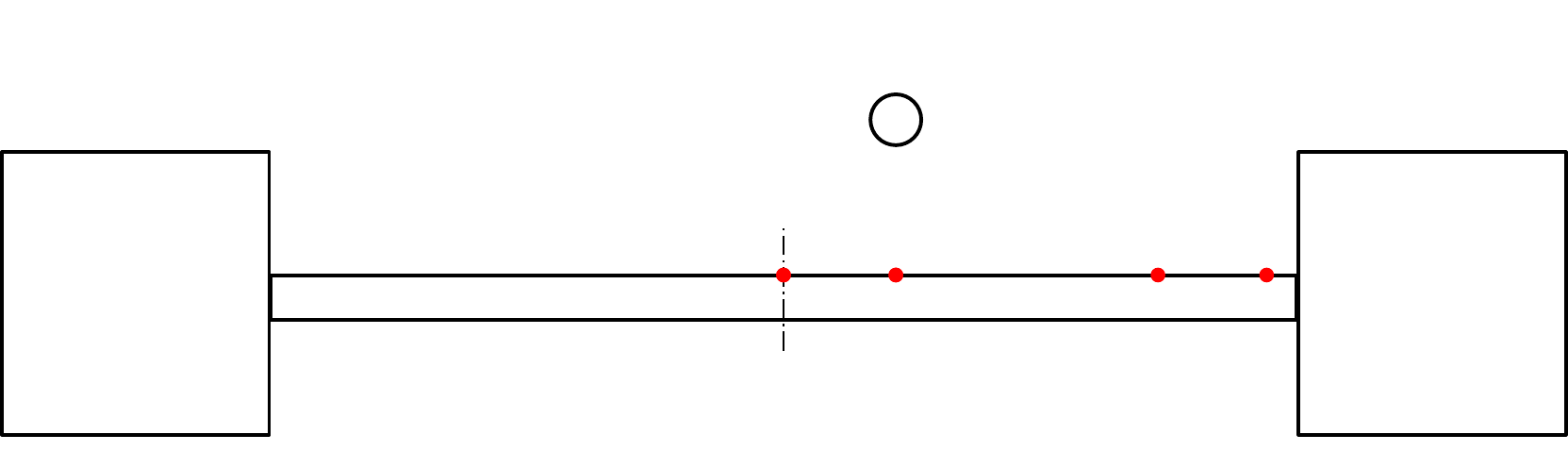}
\caption{
Velocity response under eccentric impact in the clamped-clamped configuration.
a) velocity of $P_3$,
b) velocity of $P_4$,
c) velocity of $P\si{sph}$,
d) illustration of problem setting.
}
		\label{fig:val_fixed}
	\end{minipage}
\end{figure}
%
\fref{val_free} and \frefo{val_fixed} show the velocity response in the free-free and the clamped-clamped configuration, respectively.
The vertical velocities are shown at several representative points along the beam as well as a point on the sphere (as labelled in \fref{val_free}e, \fref{val_free}d).
{
The impact process can be inferred from the time evolution of the velocity of the point $P\si{sph}$ on the sphere:
The sphere drops as a rigid body and has a velocity of $1,100~\mathrm{mm/s}$ when it comes into contact with the beam.
The velocity of the sphere changes rapidly due to the contact force until it re-bounces after about $40~\mu\mathrm{s}$ (contact duration).
}
The depicted time span of $500~\mu\mrm{s}$, hence, not only includes the impact but also a certain section of the response after the impact.
{
The depicted time span is sufficiently short that the sphere does not impact again.
}
The results of repeated tests are shown as spread.
For clarity, again, explicit FEA results are depicted only for the free-free configuration (\fref{val_free}) and implicit FEA results only for the clamped-clamped configuration (\fref{val_fixed}).
It should be remarked that the results of implicit and explicit FEA are generally very similar.
\\
As can be clearly seen by the narrow variability spread, the repeatability is excellent in both configurations.
Thus, the design of the test rig can be viewed as successful in this regard.
Overall, the computational predictions agree very well with the measurements.
The results of the clamped-clamped configuration are matching well, but not as well as in the free-free configuration.
This is attributed to larger uncertainties in the clamped-clamped configuration, especially due to the modeling of the adhesive between beam and clamping blocks.
\\
{
The velocity of the sphere is determined at point $P\si{sph}$ both in the measurement and in the simulation.
}
As mentioned before, the elastic modes of the sphere start from $\ze{230}{kHz}$,
{
which is much higher than the highest relevant vibration frequency, and thus the sphere deforms quasi-statically.
Most of the elastic, quasi-static deformation occurs within the immediate vicinity of the contact point.
}
Thus, the sphere's velocity at point $P\si{sph}$ {is essentially its} rigid body velocity.
Minor prediction errors ($\approx 3\%$) of the sphere's rebound velocity can be observed.
These are attributed to the plastic deformation occurring during the first few impact events prior to the depicted measurements.
As explained above, the initial plastic deformation leads to the formation of a shallow crater on the beam with locally increased stiffness.
In this sense, the \emph{impact} becomes \emph{harder}.
These effects are not accounted for in the model.
Due to the harder impact, more energy is transferred into waves and the sphere's rebound velocity is smaller in magnitude as in the simulation.

\subsubsection{Force pulse}\label{sec:forcePulse}
\begin{figure}[h!]
	\begin{minipage}{0.49\textwidth}
		a)\vfill
		\includegraphics[width=\textwidth]{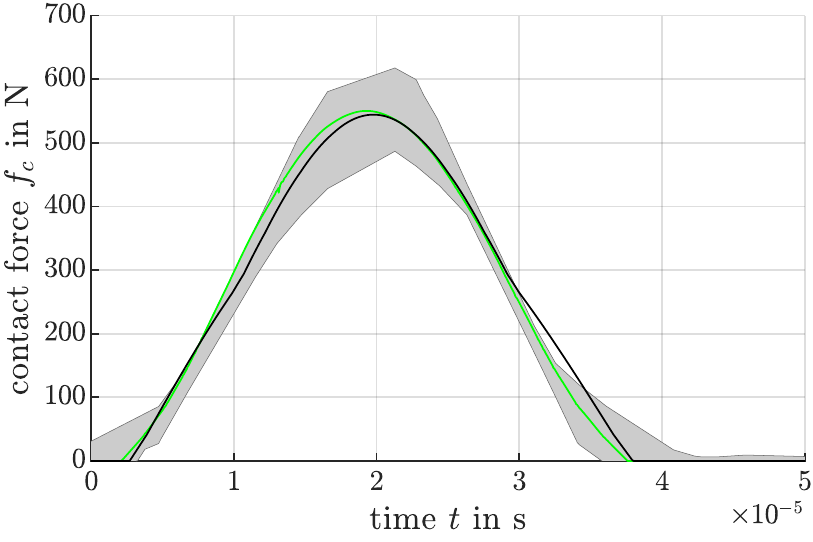}
	\end{minipage}
	\hfill
	\begin{minipage}{0.49\textwidth}
		b)\vfill
		\includegraphics[width=\textwidth]{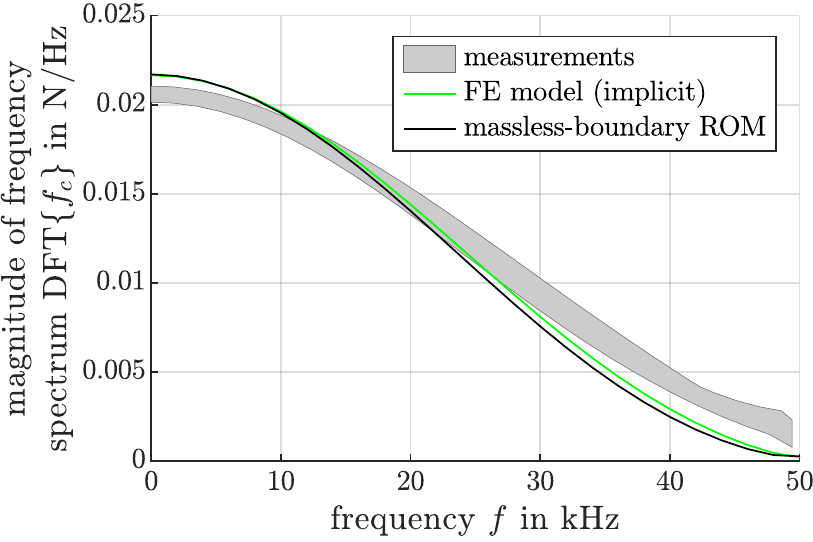}
	\end{minipage} \\
	\begin{minipage}{0.49\textwidth}
		c)\vfill
		\includegraphics[width=\textwidth]{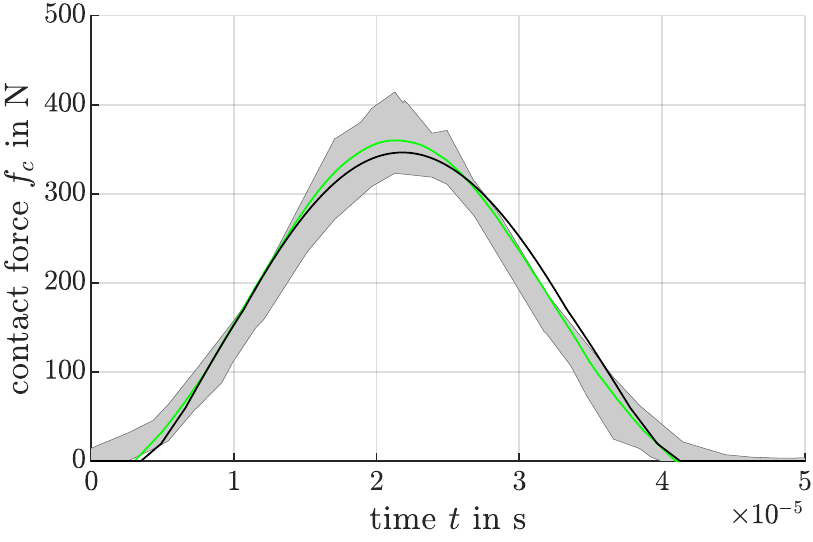}
	\end{minipage}
	\hfill
	\begin{minipage}{0.49\textwidth}
			d)\vfill
			\includegraphics[width=\textwidth]{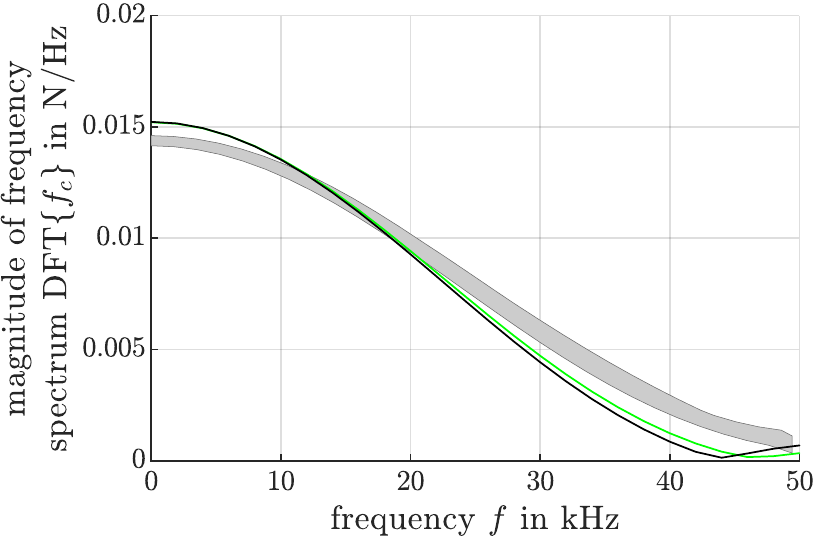}
	\end{minipage}\\
\caption{
Force pulse and its Fourier transform:
(a) and (c) force pulse $f_{\mrm c}(t)$,
(b) and (d) Fourier transform of $f_{\mrm c}(t)$;
(a)-(b) are in free-free configuration,
(c)-(d) are in clamped-clamped configuration.
}\label{fig:forcePulse}
\end{figure}
\fref{forcePulse} shows the force pulse, \ie, the contact force $f_{\mrm c}$ vs. time $t$, and its discrete Fourier transform.
In the model, $f_{\mrm c}$ is obtained as integral measure over all nodes in contact.
In the experiment, the contact force is obtained from Newton's second law, $f_{\mrm c} = m_{\mrm{sph}}\dd v_{\mrm{sph}}/\dd t$ using the known mass of the sphere,  $m_{\mrm{sph}}$, and numerical differentiation of the measured velocity, $v_{\mrm{sph}}$. 
{
Note that the sphere velocity is measured at point $P_{\mrm{sph}}$, and it is here treated as rigid-body velocity as reasoned above.
}
The computational predictions are largely within the measurement spread.
Recall that the sampling rate was limited to \ze{102.4}{kHz} by the equipment.
This leads to a time resolution of about $10~\mu\mathrm{s}$ in the measurements.
Taking into consideration that the contact duration is only about $40~\mu\mathrm{s}$, this leads to a relatively coarse time evolution of the experimentally determined force.
As explained above, the real impact is harder.
Accordingly, the frequency spectrum of the measured contact force is slightly broader.
One also expects a shorter and higher force pulse.
However, the difference appears to be so small that it cannot be clearly seen with the given time resolution.

\subsubsection{Modal energy distribution and modal restitution coefficients}\label{sec:modalEnergyDistribution}
%
%
\begin{figure}[h!]
	\begin{minipage}{0.49\textwidth}
		a)\vfill
		\includegraphics[width=\textwidth]{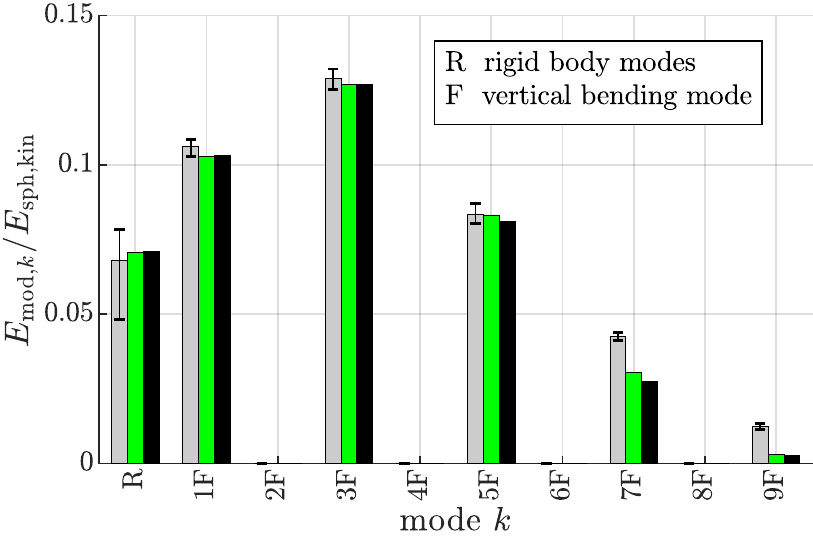}
	\end{minipage}
	\hfill
	\begin{minipage}{0.49\textwidth}
		b)\vfill
		\includegraphics[width=\textwidth]{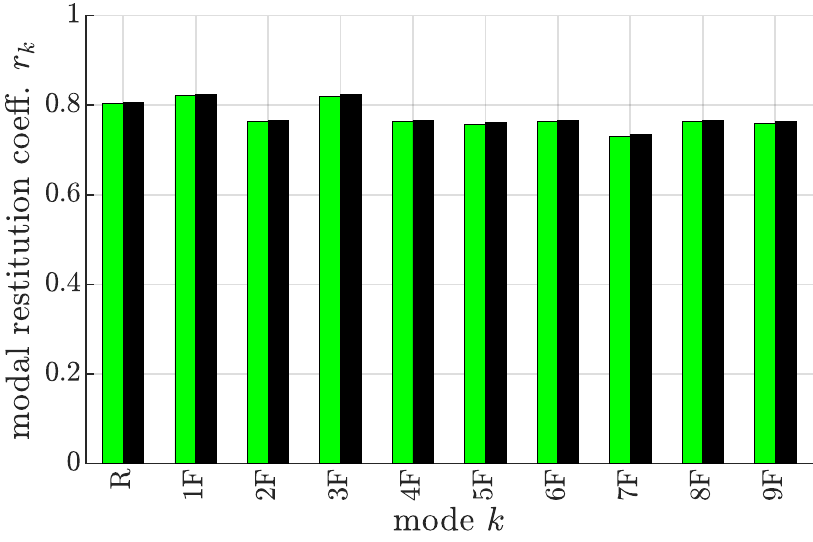}
	\end{minipage} \\
	\begin{minipage}{0.49\textwidth}
		c)\vfill
		\includegraphics[width=\textwidth]{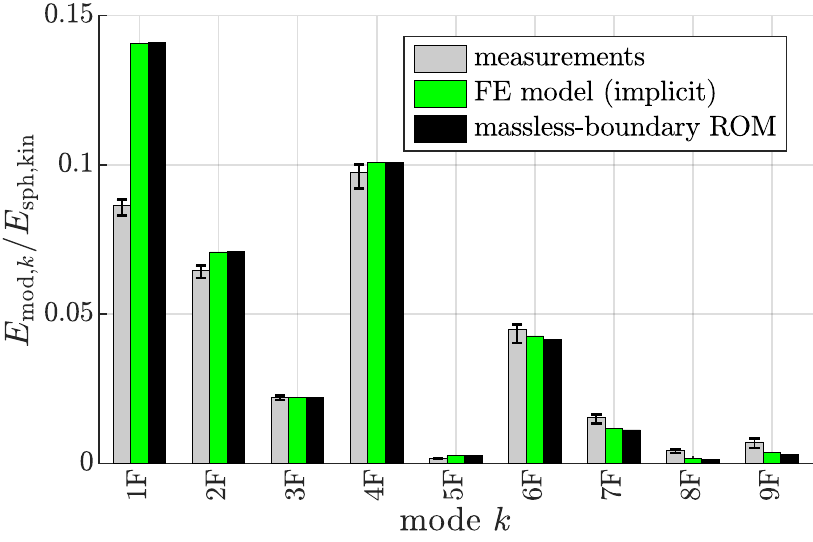}
	\end{minipage}
	\hfill
	\begin{minipage}{0.49\textwidth}
			d)\vfill
			\includegraphics[width=\textwidth]{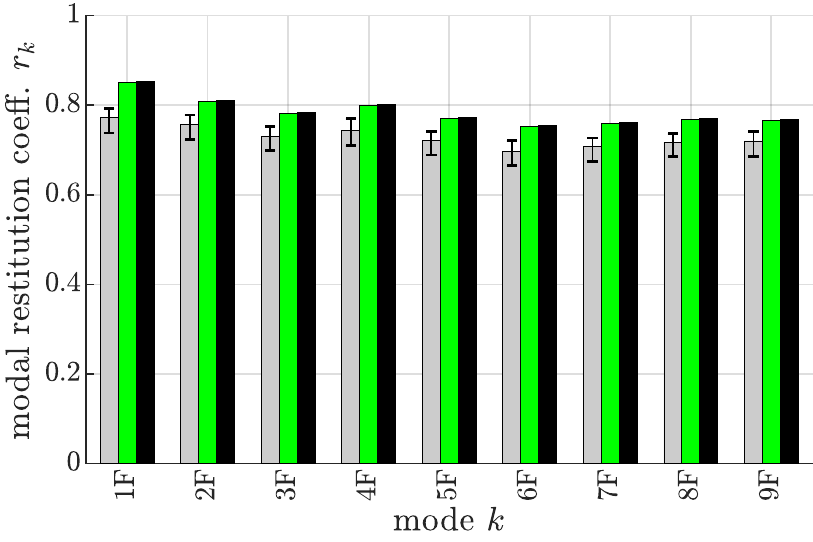}
	\end{minipage}\\
\caption{Modal energy quantities:
(a) and (c) modal energy distribution after the impact
(b) and (d) modal coefficients of restitution;
(a)-(b) are in free-free configuration,
(c)-(d) are in clamped-clamped configuration.
}\label{fig:modalEnergies}
\end{figure}
%
{
As stated in the introduction, an important technical motivation for the present work is the Impact Energy Scatterer.
Its working principle relies on transferring energy among the modes.
Under resonant excitation of a particular mode, for instance, the goal is to irreversibly transfer as much energy as possible to the off-resonant modes.
To assess the suitability of the proposed computational approach for the design of Impact Energy Scatterers, it is thus crucial to analyze the modal energy transfer.
In this section, this modal energy transfer is characterized in terms of the post-impact modal energy distribution and the so-called modal coefficients of restitution defined below.
}
\\
The modal energy, $E_{\mrm{mod},k}$ of a given mode $k$ is
\e{
E_{\mrm{mod},k} = \frac{1}{2}\left( \dot{\eta}_k^2 + \omega_k^2\eta_k^2 \right)\fk
}{Emod}
where $\eta_k$ is the $k$-th modal coordinate, $\omega_k$ is the corresponding angular natural frequency, and $\dot\square$ denotes derivative with respect to time $t$.
Here, it is presumed that the modal coordinates correspond to mass-normalized modal deflection shapes.
The modal coefficient of restitution, $r_k$, associated with mode $k$ is defined as \cite{Theurich.2021}
\e{
r_k = - \frac{v_{\mrm{sph}}\left(t_{\mrm{E}}\right) - \varphi_k\left(P_{\mathrm{impact}}\right)\,\dot\eta_k\left(t_{\mrm{E}}\right)}{v_{\mrm{sph}}\left(t_{\mrm{S}}\right)}\fk
}{mCOR}
where $t_{\mrm{S}}$ and $t_{\mrm{E}}$ denote the start and end time instants of the impact, and $\varphi_k\left(P_{\mathrm{impact}}\right)$ is the mass-normalized mode shape at the impact point with $P_{\mrm{impact}}=P_2$ in the clamped-clamped configuration and $P_{\mrm{impact}}=P_4$ in the free-free configuration (\tref{configs}).
The sphere velocity, $v_{\mrm{sph}}$ (again, treated as rigid-body velocity), and the modal deflection shape are in the vertical direction.
The modal coefficient of restitution plays an important role in the predictive design of \IESs against resonances under harmonic forcing.
If the $k$-th mode is driven near a well-separated resonance, and the \IES undergoes the typical two symmetric impacts per excitation period, then $r_k$ quantifies the loss of the $k$-th mode's energy by each impact \cite{Theurich.2021}.
This energy is irreversibly transferred to the off-resonant modes.
\\
We focus the following analysis on the vertical bending modes as listed in \tref{f_and_D}.
The modal frequencies, $\omega_k$, and the sphere velocity, $v_{\mrm{sph}}$, at the start and the end of the impact, is readily available from both experiment and computation.
To evaluate \erefs{Emod} and \erefo{mCOR}, we still need to determine the modal coordinates, $\eta_k$, their time derivatives, $\dot\eta_k$, and the mass-normalized mode shape at the impact location, $\varphi_k\left(P_{\mathrm{impact}}\right)$.
These quantities are rather straight-forward to determine from the model.
To obtain the modal coordinates, we use the relation $\mm \eta = \mm\Phi^{\mathrm T}\mm M\mm q$,
where $\mm q$ is the vector of generalized coordinates (nodal degrees of freedom in the FE model, coordinates of the component modes in the ROM),
$\mm \eta$ is the vector of modal coordinates, $\mm\Phi$ is the matrix containing the corresponding mass-normalized mode shapes as columns, and $\mm M$ is the mass matrix. 
It is more tricky to determine the modal coordinates from the measurements.
As the frequency response function was obtained only for forcing applied to point $P_2^*$ and response at point $P_2$, mass-normalized mode shapes are only available for point $P_2$, and due to symmetry also at $P_2^*$, but not $P_4$.
Consequently, different strategies for the estimation of the modal coordinates were followed in the case of the two different impact locations, $P_2$ and $P_4$.
These are described in the following two paragraphs.
Both strategies have in common that they rely only on experimental data and do not resort to the FE model.
\\
In the free-free configuration, the impact point was $P_4$.
Velocity measurements are available for $P_2$.
After the impact, the beam is assumed to respond in a linear way.
Thus, the velocity at $P_2$, $v_{P_2}$, can be expressed as
\e{
v_{P_2}(t) = v_{\mrm{RgB}} + \sum_k \omega_k\varphi_k\left(P_2\right)\,\left[~-\hat\eta_k^{\mrm c} \sin\left(\omega_k t\right) + \hat\eta_k^{\mrm s} \cos\left(\omega_k t\right)~\right]\fk
}{modalCoordinatesFreeFree}
where $v_{\mrm{RgB}}$ is the rigid body component of the velocity.
Here, it is exploited that the beam is very lightly damped in the free-free configuration, so that damped and undamped natural frequencies are practically identical and the exponential decay is negligible in the given time frame.
Indeed, the considered mode with highest frequency, the $9$-th order bending mode (9F), undergoes the strongest decay, which still amounts to only $\ee^{-D_k\omega_k t}>0.99$ after the analyzed time span of $500~\mu\mrm{s}$ (\fref{val_free}).
The rigid body velocity $v_{\mrm{RgB}}$ and the coefficients of the modal contributions, $\hat\eta_k^{\mrm c}$, $\hat\eta_k^{\mrm s}$, are determined by least-squares fitting of \eref{modalCoordinatesFreeFree} to the measured velocity $v_{P_2}$.
Here, the experimentally identified mass-normalized mode shapes at point $P_2$, $\varphi_k\left(P_2\right)$, and the natural frequencies, $\omega_k$, were used.
More specifically, to estimate $\varphi_k\left(P_2\right)$ based on the frequency response function in \eref{H1}, the symmetry between drive point $P_2^*$ and response point $P_2$ was exploited.
The components of the mass-normalized mode shapes at the impact location, $P_4$, were not determined during the experimental modal analysis.
As these are required in \eref{mCOR}, the modal restitution coefficients, $r_k$, could not be evaluated on the basis of the available experimental data for the free-free configuration.
\\
In the clamped-clamped configuration, the impact point was $P_2$.
As the mass-normalized modal deflection shapes are not available at any of the two measurement points on the beam, $P_3$ and $P_4$, least-squares fitting of \eref{modalCoordinatesFreeFree}, as in the free-free case, could not be used.
Instead, the post-impact modal response is obtained from the experimentally identified contact force $f_{\mathrm c}(t)$ (\fref{forcePulse}) using Duhamel's integral,
\ea{
\eta_k(t_{\mrm E}) &=& \frac{\varphi_k\left(P_2\right)}{\omega_k}\,\int\limits_{t_{\mrm S}}^{t_{\mrm E}} f_{\mrm c}\left(\tau\right)\, \sin\left(\omega_k\left(t_{\mrm E}-\tau\right)\right)\dd\tau\fk \label{eq:modalCoordinatesClampedClampedOne}\\
\dot \eta_k(t_{\mrm E}) &=& \varphi_k\left(P_2\right)\,\int\limits_{t_{\mrm S}}^{t_{\mrm E}} f_{\mrm c}\left(\tau\right)\, \cos\left(\omega_k\left(t_{\mrm E}-\tau\right)\right)\dd\tau\fp \label{eq:modalCoordinatesClampedClampedTwo}
}
Here, it is again exploited that damped and undamped natural frequencies are practically identical and the exponential decay term can be neglected.
In contrast to the strategy outlined in the previous section, the relevant time span is that where $f_{\mrm c}\neq 0$, \ie, the contact duration.
The strongest exponential decay is that of the 9F mode, which amounts to only $\ee^{-D_k\omega_k t}>0.994$ during the contact duration of about $40~\mu\mrm{s}$.
For the given point, $P_2$, the mass-normalized modal deflection shape is available from the experiment.
Thus, the modal restitution coefficients can be determined experimentally by substituting the post-impact modal velocities from \eref{modalCoordinatesClampedClampedTwo} into \eref{mCOR}.
\\
The modal energy distribution is shown in \fref{modalEnergies} for both configurations, normalized by the sphere's pre-impact kinetic energy, $E_{\mrm{Sph,kin}}$.
Only the bending modes in the impact direction are significantly excited.
The remaining modes, including the clamping-dominated modes in the clamped-clamped configuration, have negligible contributions and are therefore not illustrated.
The highest significantly excited natural frequency is that of the 9F mode in both configurations, which is in the range of $35~\mrm{kHz}$.
In the free-free configuration, the impact point, $P_4$, is located at the axial center of the beam.
This is the vibration node of the even-ordered modes and, thus, only the odd-ordered modes are significantly excited.
The computational methods yield almost identical results and, overall, the predictions agree well with the measurements.
In the free-free configuration, the contributions of the higher-frequency modes, 7F and 9F, are underpredicted.
This is consistent with the harder impact in the experiment.
In the clamped-clamped configuration, the contribution of the fundamental bending mode, 1F, is overpredicted.
This discrepancy is related to the deviation of the mode shape between simulation and experiment.
The corresponding mass-normalized mode shape is by a factor of about $1.19$ higher in the model, as discussed in \sref{linearFRF}.
This factor enters quadratically into the energy, and thus explains the largest part of the discrepancy.
\\
The modal coefficients of restitution are also shown in \fref{modalEnergies}.
Again, the computational methods are in almost perfect agreement.
As explained above, the restitution coefficients were only obtained experimentally in the clamped-clamped configuration.
Here, the experimental results agree very well with the simulation.
All experimentally obtained restitution coefficients are lower, which is consistent with the harder impact in the experiment.

\subsubsection{Comparative assessment of computational methods}
The computational effort of the methods should generally be viewed in the light of their accuracy.
An important aspect that affects both effort and accuracy is the contact discretization.
The largest actual contact area is obtained at the time instant of maximum compression of the contact region.
At this time instant, 13 (16) nodes of the FE model are in contact for the lower (higher) impact velocity.
Due to the described constraining of the mid-side nodes, the contact discretization is coarser in the case of the massless-boundary ROM, leading to at most only 4 (8) nodes in contact.
In spite of the relatively coarse spatial discretization, the FEA results agree very well with the measurements.
It is also remarkable that the even coarser massless-boundary ROM provides equally high prediction quality.
The good agreement holds with respect to all analyzed results, \ie, velocity response, force pulse, modal energy distribution and modal restitution coefficients.
A finer discretization would most certainly be needed, both in the FE model and the ROM, to accurately predict the actual contact area and the contact stress distribution.
\begin{table}[h!]
	\centering
	\caption{Computational effort of different simulation methods.}
	\begin{tabular}{|c|c|c|c|c|}
		\hline
		 Configuration & method & \# DOFs & \# time steps & CPU time in s \\
		\hline \hline
		\multirow{3}{*}{free-free} & \impl & $66,060$ & $5,048$ & $14,054$  \\
		\cline{2-5}
		& \expl & $66,060$ & $76,025$ & $1,037$  \\
		\cline{2-5}
		& \mlcms & 141 & $5,000$ & 1.88 \\
		\cline{2-5}
		\hline \hline
		\multirow{2}{*}{clamped-clamped} & \impl & $195,963$ & $5,035$ & $42,471$  \\
		\cline{2-5}
		& \mlcms & 187 & $5,000$ & 2.81 \\
		\hline		
	\end{tabular}
	\label{tab:performance}
\end{table}
%
\\
We now analyze the computational effort required to simulate the impact and the subsequent time span until $t=500~\mu\mrm{s}$ (as illustrated in \frefs{val_free} and \frefo{val_fixed}).
Important measures for the computational effort are listed in \tref{performance}.
{
The computations were carried out on a cluster with Intel Xeon Gold/E5 CPUs (2.6-4.4 GHz) and DDR4 RAM (2.1-2.9 GHz).
The simulation of the \mlcms was done using \MATLAB with 2 CPUs available having 16 cores each.
The simulation of the FE models was done using \ABAQUS.
In the explicit variant, 8 CPUs were used, whereas 2 CPUs and 1 GPU were used in the implicit variant.
The computation times are reported and discussed in terms of CPU times.
Thus, the part of the computation effort associated with the GPU (\ABAQUS implicit only) is not accounted for.
}
\\
In the case of the FE model, the number of degrees-of-freedom (DOFs) corresponds to the total number of nodal degrees of freedom.
In the case of the massless-boundary ROM, it corresponds to the number of degrees of freedom at the contact boundary, plus the number of retained normal modes.
This leads to a reduction of the number of degrees of freedom by 3 orders of magnitude, more specifically, a factor of about $500$ ($1,000$) in the free-free (clamped-clamped) configuration.
Consequently, the effort for the linear algebra operations involved in the computation should be decreased.
It should be noted that the segment-to-segment contact formulation in the fully implicit FEA involves the discretization of the Lagrange multiplier field, so that the actual problem dimension exceeds the number of DOFs.
It must also be emphasized that the matrices involved in the (full-order) FEA are generally sparse, while they are dense in the case of the ROM.
Thus, different algorithms are used for the linear algebra operations, so that the computational complexity scales differently with the problem dimension.
\\
In explicit schemes, the largest time step is driven by the highest relevant frequency.
In the FE model, the highest relevant frequency is proportional to the highest wave speed divided by the smallest element length.
Thus, the finer the mesh, the higher the relevant frequency, the smaller the feasible time steps.
With the massless-boundary ROM, the highest relevant frequency is that of the highest-frequency normal mode
This is an important property of the proposed massless-boundary ROM, which is not shared by most component mode synthesis methods.
Indeed, the by far most popular component mode synthesis methods, the Craig-Bampton method and the Rubin method, associate mass with the static component modes (constraint modes and residual attachment modes, respectively).
This leads to very high natural frequencies.
In contrast, the proposed MacNeal method neglects the mass associated with the residual attachment modes.
This permits the described quasi-static treatment of the contact sub-problem.
Hence, the highest relevant frequency in the dynamic simulation is that of the highest-frequency normal mode.
This frequency can be controlled by choosing the modal truncation order.
The time step sizes used in the semi-explicit integration of the ROM are very similar to those in the fully implicit FEA.
As discussed in \sref{ROMsimulation}, increasing the selected time step by up to a factor of $2.5$ does not significantly affect the quality of the results.
\\
The import of the mass and stiffness matrices of the FE model into \MATLAB, and the construction of the ROM took about \ze{310}{s} (\ze{1,360}{s}) CPU time in the free-free (clamped-clamped) configuration.
Once the ROM is constructed, the simulation of the massless-boundary ROM is 4 orders of magnitude quicker (in terms of CPU time) than the implicit FEA, more specifically by a factor of about $7,500$ ($15,000$) in the free-free (clamped-clamped) configuration.
The simulation of the massless-boundary ROM is 3 orders of magnitude quicker than the explicit FEA, more specifically by a factor of about $550$ (free-free configuration).
Therefore, the computational overhead generated by setting up the ROM rapidly pays off, especially when more impact scenarios (\eg with different initial velocities) are to be simulated.

\section{Summary and conclusions} \label{sec:conclusions}
{
We designed a test rig to analyze a sphere impacting on a free-free and a clamped-clamped beam.
In the contact region on the beam, plastic deformation takes place during the first few impacts, after which the measurements show excellent repeatability.
Three computational prediction approaches were presented.
All of these rely only on the geometry, the elastic material properties and the material density.
The first two approaches are implicit and explicit FEA, carried out using a state-of-the-art commercial tool.
The third and last approach is a recently developed massless-boundary component mode synthesis technique in conjunction with a semi-explicit integration scheme.
The computational predictions were confronted with measurements.
Based on the very good agreement in terms of linear natural frequencies and frequency response function, we conclude that the FE model and the reduced-order model represent the linear behavior of the impacted structure up to the highest frequency relevant for the impact process.
The nonlinear impact process and subsequent dynamic behavior was characterized in terms of velocity response at multiple locations, the contact force pulse, the post-impact modal energy distribution and the modal coefficients of restitution.
Throughout the results, the predictions were almost always within the repeatability spread of the measurements.
Slight discrepancies were attributed largely to the plastic deformation, not taken into account in the predictions.
Due to the very good agreement with the measurements, we conclude that the computational methods are valid and can be used, in particular, for the predictive design of \IESs.
Compared to the state-of-the-art FEA, the recently developed approach reduces the computational effort by 3 to 4 orders of magnitude, without compromising the high prediction quality.
}
\section*{Acknowledgments}
The authors are grateful to MTU Aero Engines AG, Germany, for giving permission to publish this work.

\appendix

\section{Definition of component modes, reduced stiffness and mass matrices according to the MacNeal method} \label{asec:MacNeal}
To make this article self-contained, the matrix of component modes and the reduced matrices obtained by the well-known MacNeal method are given in the following.\\
In the MacNeal method, the matrix of component modes is
\e{
\mm R =
\matrix{cc}{
		\eye                                                                 & \mm 0 \\
		\mm F_{\mathrm{ib}}^\prime\left(\mm F_{\mathrm{bb}}^\prime\right)\inv &
		\mm\Phi_{\mathrm i} - \mm F_{\mathrm{ib}}^\prime\left(\mm F_{\mathrm{bb}}^\prime\right)\inv\mm\Phi_{\mathrm b}
	}\fp
}{MacNealR}
Herein, $\eye$ denotes the identity matrix.
The different terms within this matrix are defined in the following.
\\
$\mm\Phi_{\mathrm b}$ and $\mm\Phi_{\mathrm i}$ are the corresponding partitions of the modal matrix $\mm \Phi = \left[\mm\phi_1,\ldots,\mm\phi_{\nmod}\right]$, where $\nmod$ is the number of retained normal modes.
The free interface normal modes $\mm\phi_k$ with associated natural frequency $\omega_k$ are determined from the eigenvalue problem
\e{
	\left(\mm K - \omega_k^2\mm M \right)\mm\phi_k =
	\mm 0\fk \quad \mm\phi\tra_k\mm M\mm\phi_k = 1\fk
}{NormalModes}
and are normalized with respect to the mass matrix $\mm M$.
\\
The flexibility matrix is denoted as $\mm F$ and is defined as the inverse of the stiffness matrix, $\mm F = \mm K\inv$.
The $j$-th column of $\mm F$ represents the static deflection due to a unit load applied at the $j$-th element of $\mm q$.
The columns of $\mm F$ corresponding to all boundary coordinates $\qqb$ are determined.
The associated upper and lower sub-matrices are denoted as $\mm F_{\mathrm{bb}}$ and $\mm F_{\mathrm{ib}}$, respectively.
The corresponding residual flexibility sub-matrices can then be expressed as
\ea{
	\vector{ \mm F_{\mathrm{bb}}^\prime \\ \mm F_{\mathrm{ib}}^\prime } =
	\vector{ \mm F_{\mathrm{bb}} \\ \mm F_{\mathrm{ib}} }
	- \mm\Phi \diag\left(\frac{1}{\omega_k^2}\right)\mm\Phi\tra_{\mathrm b} \fp
}
The reduced mass and stiffness matrices are
\ea{
	\Mtil = \matrix{cc}{\mm 0 & \mm 0 \\ \mathrm{sym.} & \eye}\fk \quad
	\Ktil = \matrix{cc}{\kbb & \kbi \\ \kbitra & \kii }\fk \label{eq:MacNealMK}
}
with
\ea{
\kbb &=& \left(\mm F_{\mathrm{bb}}^\prime\right)\inv\fk \label{eq:MacNealKbb}\\
\kbi &=& - \left(\mm F_{\mathrm{bb}}^\prime\right)\inv \mm\Phi_{\mathrm b}\fk \label{eq:MacNealKbi}\\
\kii &=& \diag\left(\omega_k^2\right) + \mm\Phi\tra_{\mathrm b} \left(\mm F_{\mathrm{bb}}^\prime\right)\inv\mm\Phi_{\mathrm b} \fp \label{eq:MacNealKii}
}

\end{document}